\documentclass[11pt]{article}
\usepackage{amsmath,amsfonts,amsthm,amssymb,epsfig,enumitem,accents}
\usepackage[margin=1.0in]{geometry}

\newtheorem{theorem}{Theorem}[section]
\newtheorem{corollary}[theorem]{Corollary}
\newtheorem{lemma}[theorem]{Lemma}

\newtheorem{definition}[theorem]{Definition}

\newtheorem{fact}[theorem]{Fact}

\def\softO#1{\widetilde{\mathcal{O}} \left( #1 \right)}
\def\bigO#1{\mathcal{O} \left( #1 \right)}

\def\prob#1#2{\mbox{\bf Pr}_{#1}\left[ #2 \right]}

\def\var#1#2{\mbox{\bf Var}_{#1}\left[ #2 \right]}
\def\cov#1#2{\mbox{\bf Cov}_{#1}\left[ #2 \right]}

\def\norm#1{\left\| #1 \right\|}
\def\inprod#1#2{\left\langle #1, #2 \right\rangle}
\def\setof#1{\left\{#1  \right\}}
\def\sizeof#1{\left|#1  \right|}

\def\setminus{-}

\renewcommand\aa{\boldsymbol{\mathit{a}}}
\newcommand\bb{\boldsymbol{\mathit{b}}}
\newcommand\cc{\boldsymbol{\mathit{c}}}

\newcommand\dd{\boldsymbol{\mathit{d}}}
\newcommand\ee{\boldsymbol{\mathit{e}}}

\newcommand\qq{\boldsymbol{\mathit{q}}}
\newcommand\rr{\boldsymbol{\mathit{r}}}
\renewcommand\ss{\boldsymbol{\mathit{s}}}

\newcommand\uu{\boldsymbol{\mathit{u}}}
\newcommand\vv{\boldsymbol{\mathit{v}}}
\newcommand\ww{\boldsymbol{\mathit{w}}}
\newcommand\xx{\boldsymbol{\mathit{x}}}

\newcommand\yy{\boldsymbol{\mathit{y}}}
\newcommand\zz{\boldsymbol{\mathit{z}}}

\def\bvec#1{{\mbox{\boldmath $#1$}}}

\def\one{\bvec{1}}

\newdimen\pIR
\newcommand\StevesR{{\rm I\kern\pIR R}}

\newcommand{\lmx}{\left[\begin{matrix}}
\newcommand{\rmx}{\end{matrix}\right]}

\newcommand{\R}{\StevesR}

\newcommand{\interior}[1]{\mathring{#1}}
\newcommand{\overstar}[1]{\accentset{*}{#1}}
\renewcommand{\hat}[1]{\underaccent{\bar}{#1}}

\newcommand{\mybox}[1]{\medskip \noindent\fbox{\parbox{\textwidth}{#1}}\medskip}
\newcommand{\comment}[1]{}

\newenvironment{myitemize}
{
    \begin{itemize}
        \setlength{\parskip}{0pt}
        \setlength{\parsep}{0pt}         
        \setlength{\itemsep}{1pt} 
}
{
    \end{itemize} 
}
\newenvironment{myenumerate}
{
    \begin{enumerate}
        \setlength{\parskip}{0pt}
        \setlength{\parsep}{0pt}         
        \setlength{\itemsep}{1pt} 
}
{
    \end{enumerate} 
}

\begin{document}

\title{Faster Lossy Generalized Flow via \\
Interior Point Algorithms\thanks{
This material is based upon work supported by the National Science Foundation under Grant Nos. CCF-0707522 and CCF-0634957.
Any opinions, findings, and conclusions or recommendations expressed in this material are those of the author(s) and do not necessarily reflect the views of the National Science Foundation.
}\\} 
 
\author{
Samuel I. Daitch\\
Department of Computer Science \\
Yale University
\and
Daniel A. Spielman\\
Program in Applied Mathematics and \\
Department of Computer Science \\
Yale University
}

 
\maketitle

\begin{abstract}
{
We present asymptotically 
  faster approximation algorithms for the generalized flow problems in which
  multipliers on edges are at most $1$.
For this lossy version of the maximum generalized flow problem, we obtain an additive $\epsilon$
  approximation of the maximum flow 
  in time $\softO{m^{3/2} \log^{2} (U/\epsilon)}$, where $m$
  is the number of edges in the graph, all capacities are integers in the range 
  $\setof{1, \dotsc , U}$, and all loss multipliers are ratios of integers in this range.
For minimum cost lossy generalized flow with costs in the range $\setof{1,\dotsc ,U}$, 
  we obtain a flow that has
  value within an additive $\epsilon$ of the maximum value and cost at most the optimal cost.
In many parameter ranges, these algorithms improve over the previously
  fastest algorithms for the generalized maximum flow problem by a
  factor of $m^{1/2}$
  and for the minimum cost generalized flow problem by a factor of
  approximately $m^{1/2}/ \epsilon^{2}$.

The algorithms work by accelerating traditional interior point algorithms by quickly solving the
  linear equations that arise in each step.
The contributions of this paper are twofold.
First, we analyze the performance of interior point algorithms with 
  approximate linear system solvers.
This analysis alone provides an algorithm for the standard
  minimum cost flow problem that runs in time 
  $\softO{m^{3/2} \log^{2}  U}$---an improvement of approximately
  $\softO{n / m^{1/2}}$ over previous algorithms.

Second, we 
  examine the linear equations that arise when using an interior point
  algorithm to solve generalized flow problems.
We observe that these belong to the family of symmetric M-matrices,
  and we then develop $\softO{m}$-time algorithms for solving
  linear systems in these matrices.
These algorithms reduce the problem of solving a linear system in a 
  symmetric M-matrix
  to that of solving $\bigO{\log n}$ linear systems in 
  symmetric diagonally-dominant matrices, which we can do in time
  $\softO{m}$ using the algorithm of Spielman and Teng.

All of our algorithms operate on numbers of bit length at most
  $\bigO{\log n U / \epsilon}$.

}
\end{abstract}
\thispagestyle{empty}
\newpage
\setcounter{page}{1}

\section{Introduction}
Interior-point algorithms are one of the most popular ways of solving
  linear programs.
These algorithms are iterative, and their complexity is dominated by
  the cost of solving a system of linear equations at each iteration.
Typical complexity analyses of interior point algorithms apply
  worst-case bounds on the running time of linear equations solvers.
However, in most applications the linear equations that arise are
  quite special and may be solved by faster algorithms.
Each family of optimization problem leads to a family of linear equations.
For example, the maximum flow and minimum cost flow problems require the solution
  of linear systems whose matrices are symmetric and diagonally-dominant.
The generalized versions of these flow problems result in symmetric M-matrices.


The generalized maximum flow problem is specified by a 
  directed graph $(V,E)$, 
  an inward capacity $c (e) > 0$ and a multiplier
  $\gamma (e) > 0$ for each edge $e$, and source and sink vertices $s$ and $t$.
For every unit flowing into edge $e$,  $\gamma (e)$ flows out.
In lossy generalized flow problems, each multiplier $\gamma (e)$
  is restricted to be at most $1$.
In the generalized maximum flow problem, one is asked to find the flow
  $f: E \rightarrow \R^{+}$ that maximizes the flow into $t$
  given an unlimited supply at $s$, subject to the capacity constraints
  on the amount of flow entering each edge.
In the generalized minimum cost flow problem, one also has a cost function $q (e) \geq 0$,
  and is asked to find the maximum flow of minimum cost (see \cite{AhujaMagnantiOrlin}).

In the following chart, we compare the complexity of our algorithms with
  the fastest algorithms of which we are aware.
The running times are given for networks
   in which all capacities and costs are positive integers less than $U$
   and every loss factor is a ratio of two integers less than $U$.
For the standard flow problems, our algorithms are exact,
  but for the generalized flow problems
  our algorithms find additive $\epsilon$ approximations, while
  the other approximation algorithms have multiplicative error $(1+\epsilon)$.
However, we note that our algorithms only require arithmetic with numbers of bit-length
  $\bigO{\log (n U / \epsilon )}$, whereas we suspect that the algorithms obtaining multiplicative
  approximations might require much longer numbers.

In the chart, $C$ refers to the value of the flow.

\begin{tabular}{|l|l|l|}
\hline
Exact algorithms & Approximation algorithms & Our algorithm\\
\hline
\textbf{Generalized Maximum Flow} & & \\
$\bigO{m^{2}(m+n\log n)\log U}$ \cite{GoldfarbJinOrlin} &
$\softO{m^{2} / \epsilon^{2}}$ \cite{FleischerWayne} &
$\softO{m^{1.5} \log^{2} (U/ \epsilon)}$
\\
$\bigO{m^{1.5} n^{2} \log(nU)}$ \cite{Vaidya89} & 
$\softO{m (m+n\log\log B)\log \epsilon^{-1}}$
& \\
& \cite{GFNR}\cite{TardosWayne}\cite{FleischerWayne} & \\
\hline
\textbf{Generalized Minimum Cost Flow} & & \\
$\bigO{m^{1.5} n^{2} \log(nU)}$ \cite{Vaidya89} & 
$\softO{m^{2} \log\log B/ \epsilon^{2}}$ \cite{FleischerWayne} &
$\softO{m^{1.5} \log^{2} (U/ \epsilon)}$
\\
\hline 
\textbf{Maximum Flow} & & \\
$\bigO{\min (n^{3/2}, m^{1/2}) m \log (n^{2}/m) \log U}$ & &
$\softO{m^{1.5} \log^{2} U}$\\
\cite{GoldbergRao}
&
&
\\
\hline 
\textbf{Minimum Cost Flow} & & \\
$\bigO{nm \log (n^{2}/m) \log (nC)}$ \cite{GoldbergTarjan} & & 
$\softO{m^{1.5} \log^{2} U}$\\
$\bigO{nm (\log \log U) \log (nC)}$ \cite{AhujaGoldbergOrlinTarjan} & & \\
$\bigO{(m \log n) (m + n \log n)}$ \cite{Orlin88} & & \\
\hline
\end{tabular}



\subsection{The solution of systems in M-matrices}\label{ssec:introM}
A symmetric matrix $M$ is  
  \textit{diagonally dominant} if each diagonal is at least the sum of the absolute values
  of the other entries in its row.
A symmetric matrix $M$ is an \textit{$M$-matrix} if there is a positive diagonal matrix $D$
  for which $D M D$ is diagonally dominant.
Spielman and Teng~\cite{SpielmanTengPrecon,SpielmanTengLinsolve} showed how to solve linear systems
  in diagonally dominant matrices to $\epsilon$ accuracy in time $\softO{m \log \epsilon^{-1}}$.
We show how to solve linear systems in $M$-matrices by first computing a diagonal matrix $D$
  for which $D M D$ is diagonally dominant, and then applying the solver of Spielman and Teng.
Our algorithm for finding the matrix $D$ applies the solver of Spielman and Teng an expected
  $\bigO{\log n}$ times.
While iterative algorithms are known that eventually produce such a diagonal matrix $D$,
  they have no satisfactory complexity analysis~\cite{Li,LiEtAl,FactorWidth}.

\subsection{Analysis of interior point methods}\label{ssec:introIPM}
In our analysis of interior-point methods, we examine the complexity of the
  short-step dual path following algorithm of Renegar~\cite{Renegar} as analyzed
  by Ye~\cite{YeBook}.
The key observations required by our complexity analysis are that 
  none of the slack variables become too small during the course of the algorithm
  and that the algorithm still works
  if one $\bigO{1/\sqrt{m}}$-approximately solves each linear system in the matrix norm
  (defined below).
Conveniently, this is the same type of approximation produced by our algorithm and that
  of Spielman and Teng.
This is a very crude level of approximation, and it means that these algorithms can be applied
  very quickly.
While other analyses of the behavior of interior point methods with inexact solvers have 
  appeared~\cite{RenegarCG}, we are unaware of any analyses that are sufficiently fine for our purposes.

This analysis is given in detail in Appendix \ref{app:intpt}.

\subsection{Outline of the paper}

In Section \ref{sec:intpt}, we describe the results
of our analysis of interior point methods using apprximate solvers.
In Section \ref{sec:genflow}, we describe the formulation of
the generalized flow problems as linear programs, and
discuss how to obtain the solutions from the output of
an interior-point algorithm.
In Section \ref{sec:mmats}, we give our algorithm for solving
linear systems in M-matrices.

\section{Interior-Point Algorithm using an Approximate Solver}\label{sec:intpt}

Our algorithm
uses numerical methods to solve a linear program
formulation of the generalized flow problems.
The fastest
interior-point methods for linear programs, such as that
of Renegar \cite{Renegar} require only $\bigO{\sqrt{n}}$ iterations
to approach the solution, where each iteration 
takes a step through
the convex polytope by solving a system of linear equations.

In this paper, we consider stepping through
the linear program using an only an approximate solver,
i.e. an algorithm $\xx = \mathtt{Solve}(M,\bb,\epsilon)$
that returns a solution satisfying
\[
\norm{\xx - M^{-1}\bb}_{M} \leq \epsilon \norm{M^{-1}\bb}_{M}
\]
where the {\bf matrix norm} $\norm{\cdot}_{M}$ is given by
$\norm{\vv}_{M} = \sqrt{\vv^{T}M\vv}$.

As mentioned above,
we have analyzed the Renegar \cite{Renegar} version of the dual path-folllowing algorithm,
along the lines of the analysis that found in \cite{YeBook}, 
but modified to account for the use of an approximate solver.

In particular, using the approximate solver
we implement an interior-point algorithm
with the following properties:

\begin{theorem}\label{thm:intptalg}
$\xx = \mathtt{InteriorPoint}(A,\bb,\cc,\lambda_{min},T,\yy^{0},\epsilon)$ takes input
that satisfies
\begin{myitemize}
\item $A$ is an $n\times m$ matrix; \\
$\bb$ is a length $n$ vector;
$\cc$ is a length $m$ vector
\item $AA^{T}$ is positive definite,
and $\lambda_{min} > 0$ is a lower bound on the eigenvalues of $AA^{T}$
\item $T>0$ is an upper bound on the 
absolute values of the coordinates in the dual linear program, i.e.

\qquad
$\norm{\yy}_{\infty} \leq T$ and $\norm{\ss}_{\infty} \leq T$

\qquad 
for all $(\yy,\ss)$ that satisfy $\ss = \cc - A^{T}\yy \geq 0$

\item initial point $\yy^{0}$ is a length $n$ vector where
$A^{T}\yy^{0} < \cc$
\item error parameter $\epsilon$ satisfies $0 < \epsilon < 1$
\end{myitemize}
and returns $\xx > 0$ such that
$\norm{A\xx - \bb} \leq \epsilon$ and $\cc^{T}\xx < z^{*} + \epsilon$.

Let us define
\begin{myitemize}
\item $U$ is the largest absolute value of any entry in $A,\bb ,\cc$
\item $s^{0}_{min}$ is the smallest entry of $\ss^{0} = \cc - A^{T}\yy^{0}$
\end{myitemize}
Then the algorithm 
makes $\bigO{\sqrt{m}\log \frac{TUm}{\lambda_{min}s^{0}_{min}\epsilon}}$
calls to the approximate solver, of the form
\[
\mathtt{Solve}\left(AS^{-2}A^{T} + \vv\vv^{T},\cdot,\epsilon'\right)
\]
where $S$ is a positive diagonal matrix with condition number
$\bigO{\frac{T^{2}Um^{2}}{\epsilon}}$, and $\vv,\epsilon'$ satisfy
\[
\log \frac{\norm{\vv}}{\epsilon'}
= 
\bigO{\log \frac{TUm}{s^{0}_{min}\epsilon}}
\]
\end{theorem}

In Appendix \ref{app:intpt}, we present a complete
description of this algorithm, with analysis
and proof of correctness.

\section{Solving Generalized Flow}\label{sec:genflow}

We consider network flows on a directed graph $(V,E)$
  with $V=[n]$, $E=\{e_{1},\dotsb,e_{m}\}$, 
  source $s\in V$ and sink $t\in V$.
  Edge $e_{j}$ goes from vertex $v_{j}$ to vertex $w_{j}$.
  and has inward capacity $c(e_{j})$,
  flow multiplier $\gamma(e_{j}) < 1$,
  and cost $q(e_{j})$.

We assume without loss of generality
that $t$ has a single in-edge, which we denote as $e_{t}$,
and no out-edges.

The generalized max-flow approximation algorithm
will produce a flow
that sends no worse than $\epsilon$ less than the maximum possible
flow to the sink.

The generalized min-cost approximation algorithm
will produce a flow 
that, in addition to being within $\epsilon$ of a maximum flow,
also has cost no greater than the minimum cost
of a maximum flow (see \cite{FleischerWayne}).

\subsection{Fixing Approximate Flows}

The interior-point algorithm described in the previous
section produces an output that may not exactly satisfy
the linear constraints $A\xx = \bb$.
In particular, when we apply the algorithm to a network flow
linear program, the output may only be an approximate flow:
\begin{definition}
An {\bf $\epsilon$-approximate flow} approximately satisfies all
capacity constraints and
flow conservation constraints. 
In particular, every edge may have flow up to $\epsilon$
over capacity, and
every vertex besides $s$ and $t$ may have
up to $\epsilon$ excess or deficit flow.

An {\bf exact flow} 
satisfies all capacity constraints and has exact flow conservation
at all vertices except $s$ and $t$.
\end{definition}

We are going to modify the graph slightly before running
the interior-point algorithm,
so that it will be easier to obtain an exact flow from
the approximate flow given by the interior-point algorithm.

Let us compute the {\em least-lossy-paths tree} $T$ rooted at $s$.
This is the tree that contains,
for each $v \in V\setminus \{s,t \}$,
the path $\pi_{s,v}$ from $s$ to $v$ that minimizes
$L(v) = \nolinebreak \prod_{e\in \pi_{s,v}} \gamma (e)^{-1}$,
the factor by which the flow along the path is diminished.
We can find this tree in time $\softO{m}$, using
Dijkstra's algorithm to solve
the single-source shortest-paths problem
with edge weights
$-\log \gamma (e)$.

Next, we delete from the graph all vertices $v$ such that
$L(v) > \frac{\epsilon}{2mnU}$.  Note that
in a maximum-flow, 
it is not possible to have
more than $\frac{\epsilon}{2n}$ flowing into such a $v$,
since at most $mU$ can flow out of $s$. 
Thus, deleting each such $v$ cannot decrease
the value of the maximum flow by more than $\frac{\epsilon}{2n}$.
In total, we may decrease the value of the maximum flow
by at most $\frac{\epsilon}{2}$.

We define $\epsilon_{FLOW}=\frac{\epsilon^{2}}{64m^{2}n^{2}U^{3}}$.
In the subsequent sections,
we show how to use the interior-point method to
obtain an 
$\epsilon_{FLOW}$-approximate flow
that has a value within $\epsilon_{4}$ of the maximum flow.
Assuming that the graph had been preprocessed as above,
we may convert the approximate flow into an exact flow:

\begin{lemma}
Suppose all vertices $v\in V\setminus \setof{s,t}$ satisfy 
$L(v) \leq \frac{\epsilon}{2mnU}$.
In $\softO{m}$ time,
we are able to convert an $\epsilon_{FLOW}$-approximate flow
that has a value within $\frac{\epsilon}{4}$ of the maximum flow
into an exact flow
that has a value within $\frac{\epsilon}{2}$ of the maximum flow.
The cost of this exact flow is no greater than the cost of the approximate flow.
\end{lemma}

\begin{proof}
Let us first fix the flows so that no vertex has more flow out than in.
We use the least-lossy-paths tree $T$,
starting at the leaves of the tree and working towards $s$.
To balance the flow at a vertex $v$ we increase the flow on the
tree edge into $v$.
After completing this process, for each $v$ we will have 
added 
a path of flow that delivers at most $\frac{\epsilon^{2}}{64m^{2}n^{2}U^{3}}$
additional units of flow to $v$.
Since $L(v) \leq \frac{\epsilon}{2mnU}$,
no such path requires more than 
$\frac{\epsilon^{2}}{64m^{2}n^{2}U^{3}}\cdot \frac{2mnU}{\epsilon} = 
\frac{\epsilon}{32mnU^{2}}$
flow on an edge,
and so in total we have added no more than $\frac{\epsilon}{32mU^{2}}$ to each edge.

Next, let us fix the flows so that no vertex has more flow
in than out.
We follow a similar procedure as above, except now we may use
any spanning tree rooted at and directed towards $t$.
Starting from the leaves, we balance the vertices by increasing flow out
the tree edge.
Since the network is lossy,
the total amount added to each edge is at most
$\frac{\epsilon^{2}}{64m^{2}n^{2}U^{3}}\cdot n \leq 
\frac{\epsilon^{2}}{64m^{2}nU^{3}}$.

Recall that we started with each edge having flow up to
$\frac{\epsilon^{2}}{64m^{2}n^{2}U^{3}}$
over capacity.
After balancing the flows at the vertices,
each edge may now be over capacity by as much as
\[
\frac{\epsilon}{32mU^{2}}
+
\frac{\epsilon^{2}}{64m^{2}nU^{3}}
+
\frac{\epsilon^{2}}{64m^{2}n^{2}U^{3}}
\leq 
\frac{\epsilon}{16mU^{2}}
\]
Since the edge capacities are at least 1,
the flow on an edge may be as much
as
$(1+\frac{\epsilon}{16mU^{2}})$ times the capacity.

Furthermore, while balancing the flows we may have added as much as
$\frac{\epsilon}{16mU^{2}}\cdot mU = \frac{\epsilon^{2}}{16U}$
to the total cost of the flow.
Assuming that the value of approximate flow was at least
$\frac{\epsilon}{4}$, its cost must also have been at least
$\frac{\epsilon}{4}$, and so we have increased the cost
by a multiplicative factor of at most $(1+\frac{\epsilon}{4U})$.

(If the approximate flow had value less than $\frac{\epsilon}{4}$,
then the empty flow trivially solves this flow rounding problem.)

By scaling the entire flow down by a multiplicative factor
of $(1+\frac{\epsilon}{4U})^{-1}$,
we solve the capacity violations, and also reduce the cost
of the exact flow
to be no greater than that of the approximate flow.
Since the value of a flow can be at most $U$,
the flow scaling decreases the value of the flow by no more than
$\epsilon/4$, as required.
\end{proof}

The above procedure produces an exact flow 
that is within $\epsilon/2$ of the maximum flow
in the preprocessed graph,
and therefore is within $\epsilon$ of the maximum flow
in the original graph.
Furthermore, the cost of the flow is no greater than 
the minimum cost of a maximum flow in the original graph.

Thus to solve a generalized flow problem,
it remains for us to describe how to
use the interior-point algorithm to generate
a $\epsilon_{FLOW}$-approximate flow
that has a value within $\epsilon/4$ of the maximum flow,
and, for the min-cost problem, also has cost no greater than the
the minimum cost of a maximum flow.

\subsection{Generalized Max-Flow}

We formulate the maximum flow problem as a linear program as follows:
Let $A$ be the $(n-2)\times m$ matrix whose 
nonzero entries are $A_{v_{j},j} = -1$ and
$A_{w_{j},j} = \gamma(e_{j})$,
but without rows corrsponding to $s$ and $t$.
Let $\cc$ be the length $m$ vector containing the edge capcities.
Let $\uu_{t}$ be the length $m$ unit vector
with a 1 entry for edge $e_{t}$.
Let the vectors $\xx_{1}$ and $\xx_{2}$ respectively denote
the flow into each edge and the unused inward capacity of each edge.  
The max-flow linear program, in canonical form, is:
\begin{align*}
& \min_{\xx_{i}} -\uu_{t}^{T}\xx_{1} 
&
\text{s.t.}\quad 
\lmx A & \\ I & I \rmx \lmx \xx_{1} \\ \xx_{2} \rmx &= \lmx 0 \\ \cc \rmx \\
&& \text{and } \quad \xx_{i}&\geq 0
\end{align*}
The constraint $A\xx_{1} = 0$ ensures that flow is conserved at every
vertex except $s$ and $t$, while the constraint
$\xx_{1} + \xx_{2} = \cc$ ensures that the capacities are obeyed.

Now, the dual of the above linear program is not bounded,
which is a problem for our interior-point algorithm.
To fix this, we modify the linear program slightly:
\begin{align*}
\min_{\xx_{i}} \left(-\uu_{t}^{T}\xx_{1} + 
\frac{4U}{\epsilon_{FLOW}}(\one_{m}^{T}\xx_{3} + \one_{n-2}^{T}\xx_{4} + \one_{n-2}^{T}\xx_{5})\right)
\qquad \text{s.t.} \quad \xx_{i} &\geq 0
\\
\text{and} \quad \lmx A & & & I & -I \\ I & I & -I & &\rmx 
\lmx \xx_{1} \\ \xx_{2} \\ \xx_{3} \\ \xx_{4} \\ \xx_{5} \rmx 
&= \lmx 0 \\ \cc \rmx 
\end{align*}
(We use $\one_{k}$ to denote the all-ones vector of length $k$.)

\begin{lemma}\label{lem:modifymaxflow}
This modified linear program has the same optimum value as the original
linear program.
\end{lemma}
\begin{proof}
Let us examine the new variables in the modified program 
and note that
$\xx_{3}$
has the effect of modifying the capacities, 
while $\xx_{4}$ and $\xx_{5}$
create excess or deficit of flow at the vertices.
Since we have a lossy network,
a unit modification of any of these values cannot change the 
value of the flow
by more than 1, and therefore must
increase the value of the modified linear program.
Thus, at the optimum we have $\xx_{3} = \xx_{4}= \xx_{5} = 0$
and so the solution is the same as that of the original linear program.
\end{proof}

The modified linear program has the following 
equivalent dual linear program:
\begin{gather*}
\max_{\yy_{i}} \cc^{T}\yy_{2} \qquad \text{s.t.} \quad 
\ss_{i} \geq 0
\\
\text{and}\quad 
\lmx A^{T} & I \\ & I \\ & -I \\ I & \\ -I & \rmx 
\lmx \yy_{1} \\ \yy_{2} \rmx 
+ \lmx \ss_{1} \\ \ss_{2} \\ \ss_{3} \\ \ss_{4} \\ \ss_{5} \rmx 
= \lmx -\uu_{t} \\ \bvec{0} \\ (4U/\epsilon_{FLOW})\cdot \one_{m} \\
(4U/\epsilon_{FLOW})\cdot \one_{n-2} \\ 
(4U/\epsilon_{FLOW})\cdot \one_{n-2} \rmx \\
\end{gather*}
\begin{lemma}\label{lem:maxflowbound}
The above dual linear program is bounded.
In particular, 
the coordinates of all feasible dual points 
have absolute value 
at most $(nU+1)\cdot \frac{4U}{\epsilon_{FLOW}}+1$.
\end{lemma}
\begin{proof}
Of the five constraints in the dual linear program,
the last four give $\frac{4U}{\epsilon_{FLOW}}$ as an 
explicit bound on the absolute value of $\yy$ coordinates.
It then follows that $\frac{8U}{\epsilon_{FLOW}}$ 
is a upper bound on the coordinates of
$\ss_{2},\ss_{3},\ss_{4},\ss_{5}$,
and the coordinates of $\ss_{1} = -\uu_{t} - A^{T}\yy_{1} - \yy_{2}$
can be at most 
$(nU+1)\cdot \frac{4U}{\epsilon_{FLOW}} + 1$.
\end{proof}

We refer to the $\ss_{i}$ variables as the slacks.
Recall that we
must provide the interior-point algorithm 
with an initial dual feasible point $\yy^{0}$ 
such that the corresponding slacks $\ss^{0}$ are bounded away from zero.
We choose the following initial
point, 
and note that the slacks are bounded from below by 
$\frac{U}{\epsilon_{FLOW}}$:
\begin{align*}
\lmx \yy^{0}_{1} \\ \yy^{0}_{2} \rmx 
&= 
\lmx \bvec{0} \\ -(2U/\epsilon_{FLOW})\cdot \one_{m} \rmx
\\
\lmx \ss^{0}_{1} \\ \ss^{0}_{2} \\ \ss^{0}_{3} \\ \ss^{0}_{4} \\ \ss^{0}_{5} \rmx 
&=
\lmx 
(2U/\epsilon_{FLOW})\cdot \one_{m} - \uu_{t} \\ 
(2U/\epsilon_{FLOW})\cdot \one_{m} \\ 
(2U/\epsilon_{FLOW})\cdot \one_{m} \\ 
(4U/\epsilon_{FLOW})\cdot \one_{n-2} \\
(4U/\epsilon_{FLOW})\cdot \one_{n-2} \rmx 
\end{align*}

We must also provide the interior-point algorithm with a lower bound
on the eigenvalues of the matrix
\[
\lmx A & & & I & -I \\ I & I & -I & &\rmx 
\lmx A^{T} & I \\ & I \\ & -I \\ I & \\ -I & \rmx
=
\lmx AA^{T} + 2I & A \\ A^{T} & 3I \rmx 
\]
Note that we may subtract $2I$ from the above matrix
and still have a positive definite matrix, so 
$\lambda_{min} = 2$ is certainly a lower bound on the eigenvalues.

Using the above values for $\yy^{0}$ and $\lambda_{min}$, 
and the bound on the dual coordinates given
in Lemma \ref{lem:maxflowbound},
we 
now call 
$\mathtt{InteriorPoint}$ on the modified max-flow
linear program, using error parameter
$\frac{\epsilon_{FLOW}}{2}$.
In the solution returned by the 
interior-point algorithm,
the vector $\xx_{1}$
assigns a flow value to each edge such that
the flow constraints are nearly satisfied: 

\begin{lemma}\label{lem:maxflowapprox}
$\xx_{1}$ is an
$\epsilon_{FLOW}$-approximate flow 
with value within $\epsilon_{FLOW}/2$
of the maximum flow.
\end{lemma}

\begin{proof}
Observe that the amount flowing into $t$
is at least $-1$ times the value of the modified linear program.
Since the interior-point algorithm generates a solution
to the modified linear program
within $\epsilon_{FLOW}/2$ of the optimum value,
which is $-1$ times the maximum flow,
the amount flowing into $t$ surely must be
within $\epsilon_{FLOW}/2$ of the maximum flow.

Now, let us note more precisely that 
the modified linear program
aims to minimize the 
objective function computed by subtracting the amount flowing
into $t$ 
from $4U/\epsilon_{FLOW}$ times the sum of the entries
of $\xx_{3}$, $\xx_{4}$, and $\xx_{5}$.
Since the minimum value of this objective function must be 
negative, and the solution returned by the interior-point algorithm
has a value within $\epsilon_{FLOW}/2$ of the minimum,
the value of this solution must be less than $\epsilon_{FLOW}/2 < U$.
The amount flowing into $t$ is also at most $U$,
so no entry of $\xx_{3},\xx_{4},\xx_{5}$
can be greater than $2U/(4U/\epsilon_{FLOW}) = \epsilon_{FLOW}/2$.

The interior-point algorithm guarantees that
\begin{align*}
\norm{A\xx_{1} + \xx_{4} - \xx_{5}} 
&< 
\frac{\epsilon_{FLOW}}{2}
&
\text{and}
&&
\norm{\xx_{1} + \xx_{2} - \xx_{3} - \cc} 
&< 
\frac{\epsilon_{FLOW}}{2}
\end{align*}
and so we may conclude that
\begin{align*}
\norm{A\xx_{1}} 
&< 
\epsilon_{FLOW}
&
\text{and}
&&
\xx_{1} \leq \cc + \epsilon_{FLOW}
\end{align*}
Indeed, this is precisely what is means for
$\xx_{1}$ to describe an
$\epsilon_{FLOW}$-approximate flow.
\end{proof}

\subsection{Generalized Min-Cost Flow}

As a first step in solving the generlized min-cost flow problem,
we solve the generalized max-flow linear program as described above,
to find a value $F$ that is within $\frac{\epsilon}{8}$
of the maximum flow.

We now formulate a linear program 
for finding the minimum cost flow that delivers $F$ units of flow to $t$:
\begin{align*}
\min_{\xx_{i}} \qq^{T}\xx_{1} \qquad 
\qquad  \text{s.t.}\quad \xx_{i}&\geq 0
\\
\text{and} \quad 
\lmx A & \\ I & I \rmx \lmx \xx_{1} \\ \xx_{2} \rmx &= 
\lmx F\cdot \ee_{t} \\ \cc \rmx \\
\end{align*}
where $\qq$ is the length $n$ vector containing the edge costs,
and $\ee_{t}$ is the length $n-1$ vector
that assigns 1 to vertex $t$ and 0 to all the other vertices
except $s$.
$A$ is the same matrix as in the max-flow linear program,
except that we include the row corresponding to $t$,
which translates to
a new constraint that $F$ units must flow into $t$.

We must again modify the linear program so that the dual will be bounded:
\begin{align*}
\min_{\xx_{i}} \left(\qq^{T}\xx_{1} + 
\left(\frac{4mU^{2}}{\epsilon_{FLOW}}\right)
\left(\one_{m}^{T}\xx_{3} + \one_{n-1}^{T}\xx_{4} + \one_{n-1}^{T}\xx_{5}\right)\right)
\qquad \text{s.t.} \quad \xx_{i}&\geq 0
\\
\text{and} \quad  
 \lmx A & & & I & -I \\ I & I & -I & &\rmx 
\lmx \xx_{1} \\ \xx_{2} \\ \xx_{3} \\ \xx_{4} \\ \xx_{5} \rmx 
&= \lmx F\cdot \ee_{t} \\ \cc \rmx 
\end{align*}
\begin{lemma}\label{lem:modifymincost}
This modified linear program has the same optimum value as the original
linear program.
\end{lemma}
\begin{proof}
We examine the new variables and note that
$\xx_{3}$ modifies the capacities, while $\xx_{4}$ and $\xx_{5}$
create excess supply (or demand) at the vertices.
A unit modification to any of these values can at best create a 
new path for one unit of flow to arrive at the sink.
This new path has cost at least 1, and it can replace 
an path in the optimum flow of cost at most $nU$,
for a net improvement in the cost of the flow of at most $nU-1$,
which is less than
$\frac{4mU^{2}}{\epsilon_{FLOW}}$.
Thus the value of the modified linear program can only increase
when these new variables are set to non-zero values.
\end{proof}
Now, the dual linear program is:
\begin{gather*}
\max_{\yy_{i}} 
\left(F\cdot\ee_{t}^{T}\yy_{1} + \cc^{T}\yy_{2}\right) 
\qquad \text{s.t.} \quad \ss_{i} \geq 0
\\
\text{and}\quad 
\lmx A^{T} & I \\ & I \\ & -I \\ I & \\ -I & \rmx 
\lmx \yy_{1} \\ \yy_{2} \rmx 
+ \lmx \ss_{1} \\ \ss_{2} \\ \ss_{3} \\ \ss_{4} \\ \ss_{5} \rmx 
= \lmx \qq \\ \bvec{0} \\ (4mU^{2}/\epsilon_{FLOW})\cdot \one_{m} \\ 
(4mU^{2}/\epsilon_{FLOW})\cdot \one_{n-1} \\ 
(4mU^{2}/\epsilon_{FLOW})\cdot \one_{n-1} \rmx
\end{gather*}

\begin{lemma}\label{lem:mincostbound}
The above dual linear program is bounded.
In particular, 
the coordinates of all feasible dual points
have absolute value at most 
$(nU+1)\cdot \frac{4mU^{2}}{\epsilon_{FLOW}}$.
\end{lemma}
\begin{proof}
Of the five constraints in the dual linear program,
the last four give $\frac{4mU^{2}}{\epsilon_{FLOW}}$ as an 
explicit bound on the absolute value of $\yy$ coordinates.
It then follows that $\frac{8mU^{2}}{\epsilon_{FLOW}}$ 
is a upper bound on the coordinates of
$\ss_{2},\ss_{3},\ss_{4},\ss_{5}$,
and the coordinates of 
$\ss_{1} = \qq - A^{T}\yy_{1} - \yy_{2}$
can be at most $(nU+1)\cdot \frac{4mU^{2}}{\epsilon_{FLOW}}$.
\end{proof}

Let us also note that
$\yy^{0} = \lmx \bvec{0} \\ -(mU^{2}/\epsilon_{FLOW})\one_{m} \rmx$
is an initial interior dual point with
all slacks at least $\frac{mU^{2}}{\epsilon_{FLOW}}$.

Using the above initial point, 
the bound on the dual coordinates from
Lemma \ref{lem:mincostbound},
and $\lambda_{min} = 2$ as in the previous section,
we run 
$\mathtt{InteriorPoint}$ on the modified min-cost
linear program, with error parameter
$\frac{\epsilon_{FLOW}}{2}$.
In the solution returned by the 
interior-point algorithm,
the vector $\xx_{1}$
assigns a flow value to each edge such that
the flow constraints are nearly satisfied:

\begin{lemma}\label{lem:mincostapprox}
$\xx_{1}$ is an
$\epsilon_{FLOW}$-approximate flow 
with value within $\frac{5\epsilon}{32}$
of the maximum flow.
\end{lemma}

\begin{proof}
Note that any flow in total cannot cost more that $mU^{2}$,
even if all edges are filled to maximum capacity.
Therefore the value of the solution output by the interior-point
algorithm can be at most 
$mU^{2} + \frac{\epsilon_{FLOW}}{2} < 2mU^{2}$,
and so in particular no entry of $\xx_{3},\xx_{4},\xx_{5}$
can be greater than $\frac{\epsilon_{FLOW}}{2}$.

Now, the interior-point algorithm guarantees that
\begin{align*}
\norm{A\xx_{1} + \xx_{4} - \xx_{5} - F\cdot \ee_{t}} 
&< 
\frac{\epsilon_{FLOW}}{2}
&
\text{and}
&&
\norm{\xx_{1} + \xx_{2} - \xx_{3} - \cc} 
&< 
\frac{\epsilon_{FLOW}}{2}
\end{align*}
and so we may conclude that
\begin{align*}
\norm{A\xx_{1} - F\cdot \ee_{t}} 
&< 
\epsilon_{FLOW}
&
\text{and}
&&
\xx_{1} 
&\leq \cc + \epsilon_{FLOW}
\end{align*}
These inequalities imply that this is a
$\epsilon_{FLOW}$-approximate flow,
and additionally that at least 
$F-\epsilon_{FLOW}$
is flowing into $t$.
Since $F$ is within $\frac{\epsilon}{8}$ of the maximum flow,
the amount flowing into $t$ must
be within 
$\frac{\epsilon}{8} + \epsilon_{FLOW} < \frac{5\epsilon}{32}$
of the maximum flow.
\end{proof}

By scaling down the $\xx_{1}$ flow slightly,
we obtain a flow that does not exceed the minimum cost
of a maximum flow:
\begin{lemma}
$\xx_{1}' = (1-\frac{\epsilon}{12U})\xx_{1}$ is an
$\epsilon_{FLOW}$-approximate flow 
with value within $\frac{\epsilon}{4}$
of the maximum flow,
and with cost at most the minimum cost
of a maximum flow.
\end{lemma}

\begin{proof}
We may assume that the value of flow $\xx_{1}$
is at least $\frac{3\epsilon}{32}$,
because otherwise the maximum flow would have to be
at most 
$\frac{3\epsilon}{32}+\frac{5\epsilon}{32}=\frac{\epsilon}{4}$,
and so the empty flow would trivially be within
$\frac{\epsilon}{4}$ of the maximum.
Therefore,
the minimum cost of a maximum flow must also at be least 
$\frac{3\epsilon}{32}$.

The interior-point algorithm guarantees
that the cost of $\xx_{1}$ does not exceed this
optimum cost by more than 
$\frac{\epsilon_{FLOW}}{2}$,
and so must also not exceed the optimum cost
by a multiplicative factor of more than
$(1+\frac{16\epsilon_{FLOW}}{3\epsilon}) < 
(1+ \frac{\epsilon}{12U})$.
Thus. 
$\xx_{1}' =  (1-\frac{\epsilon}{12U})\xx_{1}$
must have cost below the optimum.

Furthermore, since the value of the flow $\xx_{1}$
can be at most $U$, scaling down by 
$(1-\frac{\epsilon}{12U})$ cannot decrease the value of the
flow by more than $\frac{\epsilon}{12}$.
Therefore, the value of the value $\xx_{1}'$
is within 
$\frac{\epsilon}{12} + \frac{5\epsilon}{32}
< \frac{\epsilon}{4}$ of the maximum.
\end{proof}

\subsection{Running Time}

The linear systems in the above linear programs
take the form
\[
\bar{A} = 
\lmx A & & & I & -I \\ I & I & -I & &\rmx
\]
so the running time of the interior-point method depends 
on our ability to approximately solve systems of the form 
$\bar{A}S^{-2}\bar{A}^{T} + \vv \vv^{T}$,
where diagonal matrix $S$ and vector $\vv$ are
as described in Theorem \ref{thm:intptalg}.
As it turns out, this is not much more difficult than solving
a linear system in $AS_{1}^{-2}A^{T}$, where
$S_{1}$ is the upper left submatrix of $S$.

The matrix $AS_{1}^{-2}A^{T}$ is a symmetric $M$-matrix.
In the next section,
we describe how to 
approximately solve systems in such matrices in expected time
$\softO{m\log \frac{\kappa}{\epsilon}}$,
where $\kappa$ is the condition number of the matrix.
We then extend this result
to solve the systems
$\bar{A}S^{-2}\bar{A}^{T}+\vv \vv^{T}$ in time
$\softO{m\log \frac{\kappa\norm{\vv}}{\epsilon}}$,
where $\kappa$ is the condition number of 
$\bar{A}S^{-2}\bar{A}^{T}$.

\begin{theorem}\label{thm:main}
Using out interior-point algorithm,
we can solve the generalized max flow and generalized min-cost flow
problems in time
$\softO{m^{3/2}\log^{2}(U/\epsilon)}$
\end{theorem}
\begin{proof}
According to Theorem \ref{thm:intptalg},
the interior-point algorithm requires 
$\bigO{\sqrt{m}\log \frac{TUm}{\lambda_{min}s^{0}_{min}\epsilon}}$
calls to the solver.

Recall that
$T$ is an bound on the coordinates of the dual linear program,
and $s^{0}_{min}$ is the smallest slack at the initial point.
Above, we gave both of these values 
to be polynomial in $\frac{mU}{\epsilon}$,
for both the max-flow and min-cost linear programs.
We also gave $\lambda_{min}=2$ as a lower bound on the eigenvalues
of $\bar{A}\bar{A}^{T}$.
Thus, the total number of solves is $\softO{\sqrt{m}\log \frac{U}{\epsilon}}$.

Again referring to Theorem \ref{thm:intptalg},
we find that the condition number of $\bar{A}S^{-2}\bar{A}^{T}$
is be polynomial in $\frac{mU}{\epsilon}$,
as is the expression $\frac{\norm{\vv}}{\epsilon}$.
We conclude that each solve takes time $\softO{m\log \frac{U}{\epsilon}}$.

The preprocessing only took time $\softO{m}$
so we obtain a total running time of $\softO{m^{3/2}\log^{2}(U/\epsilon)}$.
\end{proof}

\subsection{Standard Min-Cost Flow}

In this section we describe how to use interior-point algorithms to
give an exact solution to the standard (i.e. no multipliers on edges)
min-cost flow problem.  

We use the following property of
the standard flow problem:

\begin{theorem}[see {\cite[Theorem 13.20]{Schrijver}}]\label{thm:integerflow}
Given a flow network with integer capacities,
and a positive integer $F$,
let $\Omega_{FLOW}$ be the set of flow vectors $\xx$ that 
flow $F$ units into $t$ and satisfy all capacity
and flow conservation constraints.  Then $\Omega_{FLOW}$ is a convex
polytope in which all vertices have integer coordinates.
\end{theorem}

Our goal is to find the flow in $\Omega_{FLOW}$ of minimum cost.
Since the cost function is linear, 
if there is a unique minimum-cost flow of value $F$,
it must occur at a vertex of $\Omega_{FLOW}$.  
By Theorem \ref{thm:integerflow} this must be an
integer flow, and we could find this flow exactly
by running the interior-point
algorithm until it is clear to which integer flow we are converging.

Unfortunately, the minimum-cost flow may not be unique.  However,
by applying the Isolation Lemma of 
Mulmuley, Vazirani, and Vazarani \cite{MVV},
we can modify the cost function slightly so that the
minimum-cost flow is unique, and is also a minimum-cost flow under the
original cost function.

Let us first state a modified version of the Isolation Lemma:

\begin{lemma}[see {\cite[Lemma 4]{KlivansSpielman}}]\label{lem:isolation}
Given any collection of linear functions on $m$ variables with integer 
cooefficients in the range $\{0,\dots ,U \}$.
If each variable is independently set uniformly at random to a 
value from the set $\{0,\dots ,2mU \}$, then with probability at least
$1/2$ there is a unique function in the collection 
that takes minumum value.
\end{lemma}

We now describe how to force the minimum-cost flow to be unique:

\begin{lemma}\label{lem:perturbcosts}
Given a flow network with capacities and costs in the set
$\setof{1,2,\dots ,U}$,
and a positive integer $F$,
modify the cost of each edge independently by adding a number 
uniformly at random from the set
$\setof{\frac{1}{4m^{2}U^{2}},\frac{2}{4m^{2}U^{2}},\dots ,
\frac{2mU}{4m^{2}U^{2}}}$.
Then with probability at least $1/2$, the modified network has
a unique minimum-cost flow of value $F$, and this flow is also
a minimum-cost flow of value $F$ in the original network.
\end{lemma}

\begin{proof}
The modified cost of a flow at a vertex of $\Omega_{FLOW}$ is a linear 
function of $m$ 
independent variables chosen uniformly at random from
the set
$\setof{\frac{1}{4m^{2}U^{2}},\frac{2}{4m^{2}U^{2}},\dots ,
\frac{2mU}{4m^{2}U^{2}}}$.
where the coefficients are the coordinates of the flow vector,
which by Lemma \ref{thm:integerflow}
are integers in the range $\{0,\dots ,U \}$.
So the Isolation Lemma tells us that with probablity at least $1/2$,
there is a unique vertex of $\Omega_{FLOW}$ with minumum modified cost.

Now, any vertex that was not originally of minimum cost
must have been more expensive than the minimum cost by an integer.
Since the sum of the flows on all edges can be at most $mU$,
and no edge had its cost increased by more than $\frac{1}{2mU}$,
the total cost of any flow cannot have increased by more than $1/2$.
Thus, a vertex that was not originally of minimum cost
cannot have minimum modified cost.
\end{proof}

We may now give an exact algorithm for standard minimum-cost flow.
Note that this algorithm works for any integer flow value, but 
in particular we may easily
find the exact max-flow value by running the interior-point max-flow
algorithm with an error of $1/2$, since we know the max-flow value
is an integer.

\begin{lemma}\label{lem:standardmincostflow}
To solve the standard minimum-cost flow problem
in expected time $\softO{m^{3/2}\log^{2}U}$,
perturb the edge costs as in Lemma \ref{lem:perturbcosts},
then run the min-cost flow interior point algorithm with an error
of $\frac{1}{12m^{2}U^{3}}$, and round the flow on each edge
to the nearest integer.
\end{lemma}

\begin{proof}
Let us prove correctness 
assuming that the modified costs do isolate a unique minimum-cost flow.
The running time then follows directly from Theorem \ref{thm:main},
and the fact from Lemma \ref{lem:perturbcosts} 
that after a constant number of tries we can expect
the modified costs to yield a unique minimum-cost flow.

We first note that the modified edge costs are
integer multiples of $\delta = \frac{1}{4m^{2}U^{2}}$.  Therefore,
by Theorem \ref{thm:integerflow} 
the cost of the minumum-cost flow is at least $\delta$ less
than the cost at any other vertex of $\Omega_{FLOW}$.

Now, the flow returned by the interior-point algorithm can be expressed
as a weighted average of the vertices of $\Omega_{FLOW}$.
Since the cost of 
this flow is within $\frac{1}{12m^{2}U^{3}} = \frac{\delta}{3U}$
of the minimum cost, this weighted average must assign a combined weight of
at most $\frac{1}{3U}$ to the non-minimum-cost vertices.
Therefore, the flow along any edge differs
by at most $1/3$ from the minimum-cost flow.
So by rounding to the nearest integer flow, we 
obtain the minimum-cost flow.
\end{proof}


\section{Solving linear systems in symmetric M-Matrices}\label{sec:mmats}

A symmetric \textit{$M$-matrix} is a positive definite symmetric 
  matrix with non-positive off-diagonals (see, e.g. \cite{HornJohnson,AxelssonBook,BermanPlemmons}).
Every $M$-matrix has a factorization of the form $M = A A^{T}$
  where each column of $A$ has at most $2$ nonzero entries~\cite{FactorWidth}.
Given such a factorization of an $M$-matrix, we we will show how to solve linear systems
  in the $M$-matrix in nearly-linear time.
Throughout this section, $M$ will be an $n \times n$ symmetric $M$-matrix and
  $A$ will be a $n\times m$ matrix with 2 nonzero entries per column such that
  $M = A A^{T}$.
Note that $M$ has $\bigO{m}$ non-zero entries.

Our algorithm will make use of the Spielman-Teng $\softO{m}$ expected time 
 approximate solver for linear systems in symmetric diagonally-dominant matrices,
  where we recall that a symmetric matrix is 
  \textit{diagonally-dominant} if each diagonal is at least the sum of the absolute values
  of the other entries in its row.
It is \textit{strictly diagonally-dominant} if each diagonal execceds each corresponding sum.

We will use the following standard facts about symmetric $M$-matrices,
which can be found, for example, in \cite{HornJohnson}:
\begin{fact}\label{fact:mmatrix}
If $M = \lmx M_{11} & M_{12} \\ M_{12}^{T} & M_{22} \rmx$ is a symmetric
$M$-matrix with $M_{11}$ a principal minor,
then:
\begin{myenumerate}
\item $M$ is invertible and $M^{-1}$ is a nonnegative matrix. 
\item $M_{12}$ is a nonpositive matrix.
\item $M_{11}$ is an $M$-matrix.
\item \label{fact:schurism}
The {\bf Schur complement} $S = M_{22} - M_{12}^{T}M_{11}^{-1}M_{12}$ is
an $M$-matrix.
\item \label{fact:schurdiag}
If all eigenvalues of $M$ fall in the range 
$[\lambda_{min},\lambda_{max}]$, then so do all diagonal entries of $S$.
\item For any positive diagonal matrix $D$, $DMD$ is an $M$-matrix.
\item \label{fact:scaling} 
There exists a positive diagonal matrix $D$ such that 
$DMD$ is strictly diagonally-dominant.
\end{myenumerate}
\end{fact}

Our algorithm will work by finding a diagonal matrix $D$ for which
  $D M D$ is diagonally-dominant, providing us with a system
  to which we may apply the solver of Spielman and Teng.
Our algorithm builds $D$ by an iterative process.
In each iteration, it decreases the number of rows that are not dominated by their diagonals
  by an expected constant factor.
The main step of each iteration involves the solution of $\bigO{\log n}$
  diagonally-dominant linear systems.
For simplicity, we first explain how our algorithm would work if we made 
  use of an algorithm
  $\xx = \mathtt{ExactSolve}(M,\bb )$ that exactly solves 
  the system $M\xx = \bb$, for diagonally-dominant $M$.
We then explain how we may substitute an approximate solver.

The key to our analysis is the following lemma, which says that if we multiply an $M$-matrix
  by a random diagonal matrix, then a constant fraction of the diagonals probably dominate their rows.  
\begin{lemma}[Random Scaling Lemma]\label{lem:randscale}
Given an $n\times n$ $M$-matrix $M$, 
and positive real values $\zeta \leq  1$ and $r \leq  \frac{1}{4}$,
let $D$ be a random diagonal $n \times n$ matrix
where each diagonal entry $d_{i}$ is chosen independently and uniformly
from the interval $(0 ,1)$.

Let $T\subset [n]$ be the set of rows of $MD$ with sums
at least $r$ times the pre-scaled diagonal, i.e.
\[
T = \{i\in [n ] : (MD\bvec{1})_{i} \geq rm_{ii}\}
\]

With probability at least $\frac{1-4r}{4r+7}$,
we have
\[
\sizeof{T} \geq 
\left(\frac{1}{8} - \frac{r}{2} \right)
\left(1-\beta-\frac{2}{3\zeta} \right)n
\]
where $\beta$ is the fraction of the diagonal entries of $M$
that are less than $\zeta$ times the average diagonal entry.
\end{lemma}
Note in particular that for $r=0$,
$T$ is the set of rows dominated by their diagonals.

We will use the Random Scaling Lemma to decrease the number of rows that are not dominated
  by their diagonals.
We will do this by preserving the rows that are dominated by their diagonals, and applying this lemma
  to the rest.
Without loss of generality we write
$M = \lmx M_{11} & M_{12} \\ M_{12}^{T} & M_{22} \rmx = 
\lmx A_{1}A_{1}^{T} & A_{1}A_{2}^{T} \\ A_{2}A_{1}^{T} & A_{2}A_{2}^{T} \rmx$,
where the rows
in the top section of $M$ are the ones that are 
already diagonally-dominant,
so in particular $M_{11}$ is diagonally-dominant.
Let $S=M_{22} - M_{12}^{T}M_{11}^{-1}M_{12}$ be the Schur complement and
 let $S_{D}$ be the matrix containing only the diagonal entries of $S$.

We construct a random diagonal matrix $D_{R}$ of the same size
as $M_{22}$ by choosing each 
diagonal element independently and uniformly from $(0,1)$.
We then create diagonal matrix $D = \lmx D_{1} & \\ & D_{2} \rmx$ where
$D_{2} = S_{D}^{-1/2}D_{R}$ 
and the diagonal entries of $D_{1}$ 
are given by $-M_{11}^{-1}M_{12}D_{2}\one$.
We know that the diagonal entries of $D_{1}$ are positive
because Fact \ref{fact:mmatrix} tells us that
$M_{11}^{-1}$ is nonnegative and $M_{12}$ is nonpositive.

We now show that the first set of rows of $DMD$ are diagonally-dominant,
  and a constant fraction of the rest probably become so as well.
Since $M$ is an $M$-matrix and $D$ is positive diagonal, 
$DMD$ has no positive off-diagonals.
Therefore, 
the diagonally-dominant rows of $DMD$ are the rows with 
nonnegative row sums.  The row sums of $DMD$ are:
\begin{align*}
DMD\one
&=
\lmx D_{1}M_{11}D_{1}\one + D_{1}M_{12}D_{2}\one \\
D_{2}M_{12}^{T}D_{1}\one + D_{2}M_{22}D_{2}\one
\rmx 
\\
&=
\lmx 0 \\
D_{2}SD_{2}\one
\rmx 
=
\lmx 0 \\
D_{R}S_{D}^{-1/2}SS_{D}^{-1/2}D_{R}\one
\rmx 
\end{align*}
Note that the diagonal entries of $S_{D}^{-1/2}SS_{D}^{-1/2}$ are all 1.
Thus by invoking Lemma \ref{lem:randscale} with 
$r=0$ and $\zeta = 1$, we find that
there is a $1/7$ probability
that at least $1/24$ of the row sums in the bottom section of $DMD$
become nonnegative.  Furthermore, we see that row sums in the top section
remain nonnegative.

The only problem with this idea is that 
in each iteration
it could take $\softO{mn}$
  time to compute the entire matrix $S$.  
Fortunately, 
we actually only need to compute the diagonals of $S$, (i.e. the matrix
$S_{D}$).
In fact, we only actually need a diagonal matrix $\Sigma$ that
approximates $S_{D}$.
As long the diagonals of $\Sigma^{-1/2}S\Sigma^{-1/2}$
fall in a relatively narrow range, we can still
use the Random Scaling Lemma to get a constant fraction
of improvement at each iteration.

To compute
these approximate diagonal values 
quickly, we use the random projection
technique of
Johnson and Lindenstrauss~\cite{JL}.
In Appendix \ref{app:mmats}, we prove the following
variant of their result,
that deals with random projections into
a space of constant dimension:
\begin{theorem}\label{lem:randproj}
For all constants $\alpha,\beta,\gamma,p\in (0,1)$, 
there is a positive integer $k = k_{JL}(\alpha,\beta,\gamma,p)$
such that the following holds:

For any vectors $\vv_{1},\dotsc ,\vv_{n}\in \R^{m}$
let $R$ be a $k \times m$ matrix with entries
chosen independently at random from the standard normal distribution,
and let $\ww_{i} = \sqrt{\frac{1}{k}}R\vv_{i}$.

With probability at least $p$ both of the following hold:
\begin{myenumerate}
\item 
$\sum_{i=1}^{n}\frac{\norm{\vv_{i}}^{2}}{\norm{\ww_{i}}^{2}}
\leq (1+\gamma)n$
\item 
$\sizeof{\setof{i: \frac{\norm{\vv_{i}}^{2}}{\norm{\ww_{i}}^{2}} < 1-\alpha}} 
\leq 
\beta n$
\end{myenumerate}
\end{theorem}
Let us note that 
\[
S = A_{2}(I - A_{1}^{T}M_{11}^{-1}A_{1})A_{2}^{T} =
A_{2}(I - A_{1}^{T}M_{11}^{-1}A_{1})^{2}A_{2}^{T}
\]
because $(I - A_{1}^{T}M_{11}^{-1}A_{1})$ is a projection matrix.
So if we let $\aa_{i}$ denote the $i$th row of $A_{2}$, we can
write the $i$th diagonal of $S$ as
$s_{ii} = \|(I - A_{1}^{T}M_{11}^{-1}A_{1})\aa_{i}^{T}\|^{2}$.
Then if we use Theorem \ref{lem:randproj} 
to create a random projection
matrix $R$,
$\|R(I - A_{1}^{T}M_{11}^{-1}A_{1})\aa_{i}^{T}\|^{2}$ gives
a good approximation to $s_{ii}$.  
Moreover, we can use one call to $\mathtt{ExactSolve}$
to compute each of the constant number of rows of the matrix 
$P=R(I - A_{1}^{T}M_{11}^{-1}A_{1})$.
Since $A_{2}$ has $\bigO{m}$ entries,
we can compute $PA_{2}^{T}$ in $\bigO{m}$ time,
and obtain all the approximations
$\|P\aa_{i}^{T}\|^{2}$ in $\bigO{m}$ time,
yielding the desired approximations of all $s_{ii}$ values.

\begin{figure}[t]
\mybox{
$\xx = \mathtt{ExactMMatrixSolve}(A,\bb)$ \\
Given: $n\times m$ matrix $A$, where $M=AA^{T}$
is an M-matrix and $A$ has at most 2 non-zeros per column.
\\
Returns: $\xx$ satisfying $M\xx = \bb$ 
\begin{myenumerate}
\item Set $D := I$.
\item Until $DMD$ is diagonally dominant do:
\begin{myitemize}
\setlength{\itemsep}{0pt}
\setlength{\parsep}{0pt}
\item [a.]
Permute so that
$DMD = \lmx D_{1}M_{11}D_{1} & D_{1}M_{12}D_{2} \\
D_{2}M_{12}^{T}D_{1} & D_{2}M_{22}D_{2} \rmx =
\lmx D_{1}A_{1}A_{1}^{T}D_{1}& D_{1}A_{1}A_{2}^{T}D_{2} \\
D_{2}A_{2}A_{1}^{T}D_{1} & D_{2}A_{2}A_{2}^{T}D_{2} \rmx
$ 
has  the diagonally dominant rows in the top section.
Let $\aa_{1},\dotsc \aa_{\nu}$ be the rows of $A_{2}$.

\item [b.] Set $k=k_{JL}(\frac{1}{100},\frac{1}{5},\frac{1}{100},\frac{1}{3})$, 
and let $R$ be a random 
$k \times m$ matrix with independent standard normal entries.
Let $\rr_{i}$ be the $i$th row of $R$.
\item [c.] For $i=1,\dotsc ,k$, compute 
$\qq_{i}^{T} = 
\mathtt{ExactSolve}(D_{1}M_{11}D_{1},D_{1}A_{1}\rr_{i}^{T})$.
\item [d.] Set $Q = \lmx \qq_{1}^{T} & \dotsb & \qq_{k}^{T}\rmx^{T}$.
\item [e.] Let $\Sigma$ be the $\nu\times \nu$ diagonal matrix with entries 
$\sigma_{i} = \|(R-QD_{1}A_{1})\aa_{i}^{T}\|^{2}$.
\item [f.] Let $D_{R}$ be a uniform random $\nu\times \nu$ diagonal matrix with
diagonal entries in $(0,1)$.
\item [g.] Set $D'_{2} = \Sigma^{-1/2}D_{R}$
\item [h.] Set $D'_{1}$ to be the matrix with diagonal
$D_{1}\cdot \mathtt{ExactSolve}(D_{1}M_{11}D_{1},-D_{1}M_{12}D'_{2}\one)$
\item [i.] Set $D := \lmx D'_{1} & \\ & D'_{2} \rmx$
\end{myitemize}
\item Return
$\xx = D\cdot 
\mathtt{ExactSolve}(DMD,D\bb)$
\end{myenumerate}
}
\caption{Algorithm for solving a linear system in a symmetric M-matrix.
To speed up the algorithm we will replace the exact solver 
with the Spielman Teng approximate solver.}
\label{fig:matrixalg}
\end{figure}

Our suggested algorithm, still using an exact solver,
is given in Figure \ref{fig:matrixalg}.
To make this algorithm fast, we replace the calls to the exact solver with
  calls to the approximate solver 
  $\mathtt{STSolve}$ of Spielman and Teng:

\begin{theorem}[Spielman-Teng~\cite{SpielmanTengPrecon,SpielmanTengLinsolve}]\label{st}
The
algorithm $\xx = \mathtt{STSolve}(M,\bb,\epsilon)$
takes as input a symmetric diagonally-dominant $n\times n$ matrix $M$
with $m$ non-zeros, 
a column vector $\bb$, and an error
parameter $\epsilon > 0$, and returns 
in expected time $\softO{m \log(1/\epsilon)}$ a column vector $\xx$ satisfying
\[
\|\xx - M^{-1}\bb\|_{M} \leq \epsilon \|M^{-1}\bb\|_{M}
\] 
\end{theorem}

We define the algorithm 
$\mathtt{MMatrixSolve}(A,\bb,\epsilon,\lambda_{min},\lambda_{max})$ 
as a modification of the algorithm
  $\mathtt{ExactMatrixSolve}$ in Figure \ref{fig:matrixalg}.
For this algorithm we need to provide
  upper and lower bounds $\lambda_{max},\lambda_{min}$
  on the eigenvalues of the matrix $A$, and the
  running time will depend on $\kappa = \lambda_{max}/\lambda_{min}$.

The modifications are that we need to set parameters:
\begin{gather*}
\delta = (1/24)\lambda_{min}^{1/2}\kappa^{-1/2}n^{-1}
\qquad 
\epsilon_{1} = .005(1.01\kappa mn)^{-1/2}
\qquad 
\epsilon_{2} = (1/72)\kappa^{-5/2}n^{-2}
\end{gather*}
and substitute the calls to \texttt{ExactSolve} in lines
$2c$, $2h$ and $3$ respectively with 
\begin{myitemize}
\item $\mathtt{STSolve}(D_{1}M_{11}D_{1},D_{1}A_{1}\rr_{i}^{T},\epsilon_{1})$
\item $\mathtt{STSolve}(D_{1}M_{11}D_{1},D_{1}(-M_{12}D'_{2}+\delta I)\one,\epsilon_{2})$ 
\item $\mathtt{STSolve}(DMD,D\bb,\epsilon)$.
\end{myitemize}

We may note that the final call to $\mathtt{STSolve}$ guarantees that
\[
\|D^{-1}\xx - D^{-1}M^{-1}\bb \|_{DMD} 
\leq 
\epsilon \|D^{-1}M^{-1}\bb \|_{DMD}
\]
or equivalently
\[
\|\xx - M^{-1}\bb \|_{M} 
\leq 
\epsilon \|M^{-1}\bb \|_{M}
\]
so the output fulfills the specification of an approximate solver,
provided that the algorithm terminates.

We can in fact bound the running time of this algorithm as follows:
\begin{theorem}
The expected running time of
the algorithm $\mathtt{MMatrixSolve}$ is
$\softO{m\log \frac{\kappa}{\epsilon}}$.
\end{theorem}

\begin{proof}
The running time is dominated by
the calls to the Spielman-Teng solver.
There are $\bigO{1}$ such solves per iterations,
each of which take time $\softO{m\log\kappa}$,
and at the conclusion of the algorithm,
there is one final call of time $\softO{m\log \epsilon^{-1}}$.

So, to prove the running time,
it suffices for us to give a $\bigO{\log m}$ bound on the 
expected number of iterations.
In particular, it suffices to show that in each iteration,
the number of non-diagonally-domainant rows 
in $DMD$ decreases by a constant fraction
with constant probability.

In analyzing
a single iteration,
we let $D = \lmx D_{1} & \\ & D_{2} \rmx$
denote the diagonal scaling at the start of the iteration,
and we let 
$D' = \lmx D'_{1} & \\ & D'_{2} \rmx$ denote the new diagonal scaling.
In Appendix \ref{app:mmats},
we prove:
\begin{lemma}\label{lem:posscaling}
$D'$ is a positive diagonal matrix.
\end{lemma}
This implies that $D'MD'$ 
has no positive off-diagonals,
thereby enabling us to check which rows of $D'MD'$ are
diagonally-dominant by looking for rows with nonnegative row sums.

We again let
$S = M_{22} - M_{12}^{T}M_{11}^{-1}M_{12}$
denote the Schur complement,
and let $S_{D}$ denote the matrix containing the diagonal entries
of $S$.
Let us also define $\tilde{S} = \Sigma^{-1/2}S\Sigma^{-1/2}$.
We know from Facts \ref{fact:mmatrix}.4 and \ref{fact:mmatrix}.6
that $\tilde{S}$ is an $M$-matrix.

Let $\tilde{S}_{D}$ be the matrix 
containing the diagonal entries of $\tilde{S}$.
In Appendix \ref{app:mmats},
we show that the row sums of $MD'$ 
are related to $\tilde{S}$ as follows:
\begin{lemma}\label{lem:rowsums}
\[
MD'\one \geq \lmx 0 \\ \Sigma^{1/2}(\tilde{S}D_{R} - \frac{1}{6}\tilde{S}_{D})\one \rmx 
\]
\end{lemma}
The upper part of the above inequality tells us
that all the row sums that were nonnegative in $DMD$
remain nonnegative in $D'MD'$.
From the lower part of the inequality and
by invoking the Random Scaling Lemma on the matrix $\tilde{S}$ with
$r=\frac{1}{6}$,
we find that with probabilty at least $\frac{1}{23}$,
the fraction of remaining rows of $D'MD'$
that now have positive row sums
is at least 
$\frac{1}{24}\left(1-\beta - 
\frac{2}{3\zeta} \right)$,
where for some $\zeta < 1$,
$\beta$ is the fraction of the diagonal entries of $\tilde{S}$
that are less than $\zeta$ times the average diagonal entry.
Indeed we prove in Appendix \ref{app:mmats}:
\begin{lemma}\label{lem:scaledschur}
With probability at least $\frac{1}{9}$,
at most $\frac{1}{5}$ of the diagonal entries of $\tilde{S}$ are smaller than 
$\left(\frac{99}{101} \right)^{3}$ times the average diagonal entry.
\end{lemma}

So with probability at least $\frac{1}{9}\cdot \frac{1}{23}$,
the fraction of rows with negative row sums in $DMD$
that now have positive row sums in $D'MD'$
is at least $\frac{1}{24}\left(1 - \frac{1}{5} - \frac{2}{3}\left(\frac{101}{99} \right)^{3} \right) > 0$. 

Thus, we may conclude that $\mathtt{MMatrixSolve}$
is expected to terminate after $\bigO{\log n}$ iterations, as claimed.
\end{proof}

\section{Final Remarks}

The reason that our interior-point algorithm currently cannot
produce an exact solution to generalized flow problems
is the dependence of our M-matrix solver on the condition
number of the matrix, even when approximating in the matrix norm.
It would be of interest to eliminate this dependence.

It would also be nice to extend the result to
networks with gains.
The main obstacle
is that the resulting linear programs may be ill-conditioned.

\bibliographystyle{alpha}
\bibliography{genflow}

\appendix

\section{Proofs for Section~\ref{sec:mmats}}\label{app:mmats}

\newtheorem*{lemma:randproj}{Lemma \ref{lem:randproj}}
\begin{lemma:randproj}
Given vectors $\vv_{1},\dotsc ,\vv_{n}\in \R^{m}$
and constants $\alpha,\beta,\gamma,p\in (0,1)$, 
for positive constant integer $k = k_{JL}(\alpha,\beta,\gamma,p)$,
let $R$ be a $k \times m$ matrix with entries
chosen independently at random from the standard normal distribution,
and let $\ww_{i} = \sqrt{\frac{1}{k}}R\vv_{i}$.

With probability at least $p$ both of the following hold:
\begin{enumerate}[label=(\roman{*})]
\item 
$\sum_{i=1}^{n}\frac{\norm{\vv_{i}}^{2}}{\norm{\ww_{i}}^{2}}
\leq (1+\gamma)n$
\item 
$\sizeof{\setof{i: \frac{\norm{\vv_{i}}^{2}}{\norm{\ww_{i}}^{2}} < 1-\alpha}} 
\leq 
\beta n$
\end{enumerate}
\end{lemma:randproj}
\begin{proof}
Let $Z_{i} = \frac{\norm{\vv_{i}}^{2}}{\norm{\ww_{i}}^{2}}$
and $Z = \sum_{i=1}^{n}Z_{i}$.

Let $\rr_{1},\dots,\rr_{k}$ be the rows of $R$,
and let $w_{ij} = k^{-1/2}\inprod{\rr_{j}}{\vv_{i}}$
be the $j$th entry of $\ww_{i}$.

Without loss of generality, we assume that all the $\vv_{i}$ are unit vectors.
Thus for any given $i$, the 
expressions $k^{1/2}w_{i1},\dots ,k^{1/2}w_{ik}$
are independent standard normal random variables.
So the expression
\[
\frac{Z_{i}}{k}
=
\frac{1}{k\norm{\ww_{i}}^{2}}
=
\frac{1}{\sum_{j=1}^{k}(\sqrt{k}w_{ij})^{2}}
\]
has inverse-chi-square distribution, with mean $\frac{1}{k-2}$
and variance $\frac{2}{(k-2)^{2}(k-4)}$.
Therefore,
$Z$
has mean $\frac{kn}{k-2}$ 
and variance at most $\frac{2k^{2}n^{2}}{(k-2)^{2}(k-4)}$,
because
\[
\var{}{Z}
=
\var{}{\sum_{i=1}^{n}Z_{i}}
=
\sum_{i,j=1}^{n}\cov{}{Z_{i}Z_{j}}
\leq 
\sum_{i,j=1}^{n}\sqrt{\var{}{Z_{i}}\var{}{Z_{j}}}
=
n^{2}\var{}{Z_{i}}
=
k^{2}n^{2}\var{}{\frac{Z_{i}}{k}}
\]
So using Cantelli's inequality,
we may conclude that 
\begin{equation}\label{eq:projprop1}
\prob{}{ Z > (1+\gamma)n }
< \frac{\var{}{Z}}{\var{}{Z}+(1+\gamma - \frac{k}{k-2})^{2}n^{2}}
\leq 
\frac{2}{2 + (k-4)(1-\frac{2}{k})^{2}(\gamma - \frac{2}{k-2})^{2}}
\end{equation}

By the same reasoning,
$\frac{k}{Z_{i}}$ has chi-square distribution,
with mean $k$ and variance $2k$.  
So using Cantelli's inequality, we find that
\[
\prob{}{
Z_{i}
<
1-\alpha
}
=
\prob{}{
\frac{k}{Z_{i}}
>
\frac{k}{1-\alpha}
}
<
\frac{\var{}{k/Z_{i}}}{\var{}{k/Z_{i}}+(\frac{k}{1-\alpha}-k)^{2}}
=
\frac{2}{2+(\frac{\alpha}{1-\alpha})^{2}k}
\]
Thus, the set $\setof{i:Z_{i} < 1- \alpha}$ has expected cardinality
less than $\frac{2}{2+(\frac{\alpha}{1-\alpha})^{2}k}n$.
So using Markov's inequality, we conclude that
\begin{equation}\label{eq:projprop2}
\prob{}{\sizeof{\setof{i:Z_{i} < 1-\alpha}} > \beta n }
<
\frac{2}{\beta(2 + (\frac{\alpha}{1-\alpha})^{2} k)}
\end{equation}
Combining inequalities \ref{eq:projprop1} and \ref{eq:projprop2} 
via the union bound, we find the probability that 
{\em (i)} and {\em (ii)} both occur is at least
\[
1 - \frac{2}{2 + (k-4)(1-\frac{2}{k})^{2}(\gamma - \frac{2}{k-2})^{2}}
- \frac{2}{\beta(2 + (\frac{\alpha}{1-\alpha})^{2} k)}
\]
which is greater than $p$ for sufficiently large $k$.
\end{proof}

\newtheorem*{lemma:randscale}{Lemma \ref{lem:randscale}}
\begin{lemma:randscale}[Random Scaling Lemma]
Given an $n\times n$ M-matrix $M$, 
and positive real values $\zeta \leq  1$ and $r \leq  \frac{1}{4}$,
let $D$ be a random diagonal $n \times n$ matrix
where each diagonal entry $d_{i}$ is chosen independently and uniformly
from the interval $(0 ,1)$.

Let $T\subset [n]$ be the set of rows of $MD$ with sums
at least $r$ times the pre-scaled diagonal, i.e.
\[
T = \{i\in [n ] : (MD\bvec{1})_{i} \geq rm_{ii}\}
\]

With probability at least $\frac{1-4r}{4r+7}$,
we have
\[
\sizeof{T} \geq 
\left(\frac{1}{8} - \frac{r}{2} \right)
\left(1-\beta-\frac{2}{3\zeta} \right)n
\]
where $\beta$ is the fraction of the diagonal entries of $M$
that are less than $\zeta$ times the average diagonal entry.
\end{lemma:randscale}

\begin{proof}
Let $M_{O}$ denote the matrix
containing only the off-diagonal elements of $M$.  
Thus, $M_{O}$ has no positive entries.

Let $B$ be the set of rows of $M$ in which the diagonal entry
is less than $\zeta$ times the average diagonal entry.  
Thus $\sizeof{B} = \beta n$.

We define a subset $J$ of rows of $M$ whose sums are not too far from
being positive.  In particular,
we let $J$ be the set of rows in which the sum of the off-diagonal
entries is
no less than $-\frac{3}{2}$ times the diagonal entry:
\[
J = 
\setof{i\in [n] : (M_{O}\bvec{1}_{n})_{i} \geq  -\frac{3}{2}m_{ii}}
\] 
Let us prove that $J$ cannot be too small.
Let $S$ be the sum of the diagonal entries of $M$.
We have:
\begin{align*}
S
&= 
\one_{n}^{T}M\one_{n}
-
\sum_{i\in [n]} (M_{O}\one_{n})_{i}
\\
&\geq 
-
\sum_{i\in [n]} (M_{O}\one_{n})_{i}
\qquad \text{(because $M$ is positive definite)}
\\
&\geq 
-\sum_{i\in [n]\setminus (J \cup B)} (M_{O}\one_{n})_{i}
\qquad \text{(because $M_{O}$ is non-positive)}
\\
&\geq 
\frac{3}{2}\sum_{i\in [n]\setminus (J \cup B)} m_{ii}
\qquad \text{(by definition of $J$)}
\\
&\geq 
\frac{3}{2}\frac{\zeta S}{n}\sizeof{[n]\setminus (J\cup B)}
\qquad \text{(by definition of $B$)}
\\
&\geq 
\frac{3}{2}\frac{\zeta S}{n}(n-\sizeof{J}-\beta n)
\end{align*}
So we see that $\sizeof{J} \geq (1-\beta - \frac{2}{3\zeta})n$

Next, let us show that the rows in $J$
have a high probability of being in $T$. 
Consider the $i$th row sum of $M_{O}D$:
\[
(M_{O}D\bvec{1})_{i} = \sum_{j\ne i}d_{j}m_{ij}
= \frac{1}{2}\sum_{j\ne i}m_{ij} + \sum_{j\ne i}(d_{j} - \frac{1}{2})m_{ij}
= \frac{1}{2}(M_{O}\bvec{1})_{i} + \sum_{j\ne i}(d_{j} - \frac{1}{2})m_{ij}
\]
Since each $(d_{j} - \frac{1}{2})$ is symmetrically distributed around zero,
we may conclude that $\frac{1}{2}(M_{O}\bvec{1})_{i}$
is the median value of $(M_{O}D\bvec{1})_{i}$.
We may also note that
$(MD\bvec{1})_{i} = (M_{O}D\bvec{1})_{i} + d_{i}m_{ii}$,
and that the values of
$(M_{O}D\bvec{1})_{i}$ and 
$d_{i}m_{ii}$ are independent.

We thus have, for $i\in J$:
\begin{align*}
\prob{}{(MD\bvec{1})_{i} \geq rm_{ii}}
&\geq  
\prob{}{(M_{O}D\bvec{1})_{i} 
\geq \frac{1}{2}(M_{O}\bvec{1})_{i} }
\cdot 
\prob{}{d_{i}m_{ii}
\geq rm_{ii}-\frac{1}{2}(M_{O}\bvec{1})_{i} }
\\
&=  
\frac{1}{2}
\cdot 
\prob{}{d_{i}m_{ii} 
\geq rm_{ii}-\frac{1}{2}(M_{O}\bvec{1})_{i}}
\\
&\geq 
\frac{1}{2}
\cdot 
\prob{}{d_{i}m_{ii} 
\geq rm_{ii}+\frac{1}{2}\cdot\frac{3}{2}m_{ii} }
\quad \text{(by definition of $J$)}
\\
&=
\frac{1}{2}
\cdot 
\prob{}{d_{i} \geq r + \frac{3}{4} }
\\
&=
\frac{1}{4} - r
\end{align*}

Thus the expected size of $J\setminus T$ is at most
$\left(r + \frac{3}{4}\right)\sizeof{J}$.
So we find

\begin{align*}
\prob{}{\sizeof{T} > 
\left(\frac{1}{8} - \frac{r}{2} \right)\sizeof{J} }
&\geq 
\prob{}{\sizeof{J\cap T} > 
\left(\frac{1}{8} - \frac{r}{2} \right)\sizeof{J} }
\\
&=
\prob{}{\sizeof{J- T} <
\left(\frac{r}{2} + \frac{7}{8} \right)\sizeof{J} }
\\
&\geq
1-\frac{r+\frac{3}{4}}{\frac{r}{2}+\frac{7}{8}}
\qquad \text{(by Markov's inequality)}
\\
&=
\frac{1-4r}{4r+7}
\end{align*}
The lemma then follows from the lower bound on $\sizeof{J}$ proven above.
\end{proof}

\newtheorem*{lemma:posscaling}{Lemma \ref{lem:posscaling}}
\begin{lemma:posscaling}
$D'$ is a positive diagonal matrix.
\end{lemma:posscaling}
\begin{proof}
$D'_{2} = \Sigma^{-1/2}D_{R}$ is trivially positive diagonal by
construction.

To check that $D'_{1}$ is positive, we use Lemma \ref{lem:aux1},
which implies that
\[
D'_{1}\one > -M_{11}^{-1}M_{12}D'_{2}\one_{n-\nu} +
\delta \left(M_{11}^{-1}\one_{n-\nu} - 
\frac{3}{4}\lambda_{max}^{-1}\one_{n-\nu} \right)
\]

To see why the above expression is positive,
recall from Fact \ref{fact:mmatrix} that $M_{11}^{-1}$ and $-M_{12}$ 
are positive matrices.  Furthermore, 
note that the diagonals of $M_{11}^{-1}$
are at least $\lambda_{max}^{-1}$.
\end{proof}

\begin{lemma}\label{lem:aux1}
\[
\norm{D'_{1}\one - 
M_{11}^{-1}(-M_{12}D'_{2}+\delta I)\one} < 
\frac{3}{4}\delta \lambda_{max}^{-1}
\]
\end{lemma}
\begin{proof}
Recall from the algorithm that
\[
D_{1}^{-1}D'_{1}\one = 
\mathtt{STSolve}(D_{1}M_{11}D_{1},
D_{1}(-M_{12}D'_{2}+\delta I)\one,\epsilon_{2})
\]
Therefore, $\mathtt{STSolve}$ guarantees that
\[
\norm{D_{1}^{-1}D'_{1}\one - 
D_{1}^{-1}M_{11}^{-1}(-M_{12}D'_{2}+\delta I)\one}_{D_{1}M_{11}D_{1}} \leq 
\epsilon_{2}
\norm{D_{1}^{-1}M_{11}^{-1}(-M_{12}D'_{2}+\delta I)\one}_{D_{1}M_{11}D_{1}}
\]
or equivalently
\[
\norm{D'_{1}\one - 
M_{11}^{-1}(-M_{12}D'_{2}+\delta I)\one}_{M_{11}} \leq 
\epsilon_{2}
\norm{M_{11}^{-1}(-M_{12}D'_{2}\one+\delta I)\one}_{M_{11}}
\]
which in turn implies that
\[
\norm{D'_{1}\one - 
M_{11}^{-1}(-M_{12}D'_{2}+\delta I)\one} \leq 
\epsilon_{2}\kappa^{1/2} 
\norm{M_{11}^{-1}(-M_{12}D'_{2}+\delta I)\one}
\]

We can then see 
\begin{align*}
\norm{D'_{1}\one - 
M_{11}^{-1}(-M_{12}D'_{2}+\delta I)\one}
&\leq 
\epsilon_{2}\kappa^{1/2} 
\norm{-M_{11}^{-1}M_{12}D'_{2}\one+\delta M_{11}^{-1}\one}
\\
&\leq
\epsilon_{2}\kappa^{1/2} 
\norm{M_{11}^{-1}M_{12}D'_{2}\one} +
\delta \epsilon_{2} \kappa^{1/2}
\norm{M_{11}^{-1}\one}
\\
&\leq
\epsilon_{2}\kappa^{1/2} 
\norm{M_{11}^{-1}M_{12}D'_{2}\one} +
\delta\epsilon_{2} \kappa^{1/2} \lambda_{min}^{-1}n^{1/2}
\\
&\leq 
\epsilon_{2}\kappa n^{1/2}
\norm{D'_{2}\one} +
\delta\epsilon_{2} \kappa^{1/2} \lambda_{min}^{-1}n^{1/2}
\qquad \text{(by Lemma \ref{lem:aux2})}
\\
&\leq
\epsilon_{2}\kappa n^{1/2}
\norm{\Sigma^{-1/2} \one} +
\delta\epsilon_{2} \kappa^{1/2} \lambda_{min}^{-1}n^{1/2}
\qquad \text{($D'_{2} < \Sigma^{-1/2}$ by construction)}
\\
&\leq
2 \epsilon_{2}\kappa n^{1/2}
\norm{S_{D}^{-1/2} \one} +
\delta\epsilon_{2} \kappa^{1/2} \lambda_{min}^{-1}n^{1/2}
\quad \text{($\Sigma^{-1/2} \leq 2S_{D}^{-1/2}$ by Lemma \ref{lem:scaledschur})}
\\
&\leq
2 \epsilon_{2}\kappa
\lambda_{min}^{-1/2} n +
\delta\epsilon_{2} \kappa^{1/2} \lambda_{min}^{-1}n^{1/2}
\qquad \text{($S_{D}^{-1/2}\one < \lambda_{min}^{-1/2}\one$ 
by Fact \ref{fact:mmatrix}.\ref{fact:schurdiag})}
\\
&= \left(
2\delta^{-1} \epsilon_{2}\kappa^{2}
\lambda_{min}^{1/2} n +
\epsilon_{2}\kappa^{1/2}n^{1/2}\right)
\delta \lambda_{max}^{-1}
\\
&= \left(
\frac{2}{3} + 
\frac{1}{72}\kappa^{-2}n^{-3/2}\right)
\delta \lambda_{max}^{-1} 
\\
&< \frac{3}{4}\delta \lambda_{max}^{-1}\
\end{align*}
\end{proof}

\newtheorem*{lemma:scaledschur}{Lemma \ref{lem:scaledschur}}
\begin{lemma:scaledschur}
With probability at least $\frac{1}{9}$,
at most $\frac{1}{5}$ of the diagonal entries of $\tilde{S}$ are smaller than 
$\left(\frac{99}{101} \right)^{3}$ times the average diagonal entry.
\end{lemma:scaledschur}

\begin{proof}
Recall that the diagonal entries of $\tilde{S}$ are 
$\tilde{s}_{ii} = \frac{s_{ii}}{\sigma_{i}}$,
where $s_{ii} = \norm{(I-A_{1}^{T}M_{11}^{-1}A_{1})\aa_{i}^{T}}^{2}$
and $\sigma_{i} = \norm{(R-QD_{1}A_{1})\aa_{i}^{T}}^{2}$.

Let us define 
$w_{i} = \frac{1}{k}\norm{R(I-A_{1}^{T}M_{11}^{-1}A_{1})\aa_{i}^{T}}^{2}$,
where 
$k = k_{JL}\left(\frac{1}{100},\frac{1}{5},\frac{1}{100},\frac{1}{3} \right)$.
By Lemma \ref{lem:randproj}, there is at least $\frac{1}{3}$ probability
that 
\begin{align*}
\frac{1}{\nu}\sum_{i=1}^{\nu} \frac{s_{ii}}{w_{i}} 
&\leq 
1.01
&\text{and}
&&
\sizeof{\setof{i:\frac{s_{ii}}{w_{i}} \leq .99}} \leq \frac{1}{5} \nu 
\end{align*}

So, by Lemma \ref{lem:sigmasclose} below, 
there is at least a $\frac{1}{3} - \frac{2}{9km} > \frac{1}{9}$ probability
that
the average diagonal entry of $\tilde{S}$ is at most
\begin{equation*}
\frac{1}{\nu}\sum_{i=1}^{\nu} \frac{s_{ii}}{\sigma_{i}}
= 
\frac{1}{\nu}\sum_{i=1}^{\nu} \frac{s_{ii}}{w_{i}}\cdot \frac{w_{i}}{\sigma_{i}}
\leq 
\frac{1.01}{k(.99)^{2}} 
\end{equation*}
and similarly we have the following bound on the number of small diagonal
entries:
\begin{align*}
\sizeof{\setof{i:\frac{s_{ii}}{\sigma_{i}} \leq  \frac{.99}{k(1.01)^{2}}}}
\leq 
\frac{1}{5} \nu 
\end{align*}

\end{proof}
\begin{lemma}\label{lem:sigmasclose}
With probability at least $1-\frac{2}{9km}$
it holds for all $i$ that
\[
\frac{1}{k(1.01)^{2}} \leq \frac{w_{i}}{\sigma_{i}} \leq \frac{1}{k(.99)^{2}}
\]
\end{lemma}
\begin{proof}
We have:
\begin{align*}
\sizeof{\sigma_{i}^{1/2} - k^{1/2}w_{i}^{1/2}}
&= 
\left|\|(R-QD_{1}A_{1})\aa_{i}^{T}\| 
- \|(R-RA_{1}^{T}M_{11}^{-1}A_{1})\aa_{i}^{T}\|
\right| \\
&\leq \|((R-QD_{1}A_{1})-(R-RA_{1}^{T}M_{11}^{-1}A_{1}))\aa_{i}^{T}\| \\
&= \|(RA_{1}^{T}M_{11}^{-1}A_{1} - QD_{1}A_{1})\aa_{i}^{T}\| \\
&\leq \|\aa_{i}\|
\sqrt{\sum_{j=1}^{k}\|\rr_{j}A_{1}^{T}M_{11}^{-1}A_{1} - \qq_{j}D_{1}A_{1}\|^{2}} \\
&= \|\aa_{i}\|
\sqrt{\sum_{j=1}^{k}\|\rr_{j}A_{1}^{T}M_{11}^{-1}D_{1}^{-1} - \qq_{j}\|_{D_{1}M_{11}D_{1}}^{2}} \\
&\leq \lambda_{max}^{1/2}
\sqrt{\sum_{j=1}^{k}\|\rr_{j}A_{1}^{T}M_{11}^{-1}D_{1}^{-1} - \qq_{j}\|_{D_{1}M_{11}D_{1}}^{2}} \\
&\text{($\norm{\aa_{i}}^{2}$ is $i$th diagonal of $M_{22}$,
so cannot exceed $M_{22}$'s largest eigenvalue)}\\
&\leq \lambda_{max}^{1/2} (\frac{s_{ii}}{\lambda_{min}})^{1/2}
\sqrt{\sum_{j=1}^{k}\|\rr_{j}A_{1}^{T}M_{11}^{-1}D_{1}^{-1} 
- \qq_{j}\|_{D_{1}M_{11}D_{1}}^{2}} 
\quad \text{(using Fact \ref{fact:mmatrix}.\ref{fact:schurdiag})}
\\
&= (\kappa s_{ii})^{1/2}
\sqrt{\sum_{j=1}^{k}\|\rr_{j}A_{1}^{T}M_{11}^{-1}D_{1}^{-1} 
- \qq_{j}\|_{D_{1}M_{11}D_{1}}^{2}} 
\\
&\leq (\kappa s_{ii})^{1/2}\epsilon_{1}
\sqrt{\sum_{j=1}^{k}\|\rr_{j}A_{1}^{T}M_{11}^{-1}D_{1}^{-1}\|_{D_{1}M_{11}D_{1}}^{2}} \quad \text{(by guarantee of $\mathtt{STSolve}$)} 
\\
&= .005\cdot  s_{ii}^{1/2}(1.01mn)^{-1/2}
\sqrt{\sum_{j=1}^{k}\|\rr_{j}A_{1}^{T}M_{11}^{-1}A_{1}\|^{2}} 
\\
&= .005\cdot  s_{ii}^{1/2}(1.01mn)^{-1/2}
\sqrt{\sum_{j=1}^{k}\|\rr_{j}\|^{2}} 
\quad \text{(because $A_{1}^{T}M_{11}A_{1}$ is a projection matrix)} 
\\
&\leq  .01 \cdot s_{ii}^{1/2}k^{1/2}(1.01n)^{-1/2} 
\\
\intertext{The above inequality does not hold with probability at most
$\frac{2}{9km}$, 
based on the fact that
expression $\sum_{j=1}^{k}\|\rr_{j}\|^{2}$
has chi-square distribution with $mk$ degrees of freedom.
}
&\leq  .01\cdot k^{1/2}w_{i}^{1/2} 
\qquad \text{(Lemma \ref{lem:randproj} implies that 
$s_{ii} \leq 1.01\cdot nw_{i}$)}
\end{align*}

So we conclude that 
\[
\sizeof{\sqrt{\frac{\sigma}{w_{i}}} - k^{1/2}} \leq .01 \cdot k^{1/2}
\]
\end{proof}

\newtheorem*{lemma:rowsums}{Lemma \ref{lem:rowsums}}
\begin{lemma:rowsums}
\[
MD'\one \geq \lmx 0 \\ \Sigma^{1/2}(\tilde{S}D_{R} - \frac{1}{6}\tilde{S}_{D})\one \rmx 
\]
\end{lemma:rowsums}

\begin{proof}
\begin{align*}
MD'\one
&=
\lmx M_{11}D'_{1}\one + M_{12}D'_{2}\one \\
M_{12}^{T}D'_{1}\one + M_{22}D'_{2}\one 
\rmx \\
&=
\lmx 0 \\
SD'_{2}\one
\rmx
+
M
\lmx D'_{1}\one + M_{11}^{-1}M_{12}D'_{2}\one 
- \delta M_{11}^{-1}\one
\\ 0 \rmx 
+
\lmx 
\delta \one \\
\delta M_{12}^{T}M_{11}^{-1}\one
\rmx
\\
&\geq 
\lmx 0 \\
SD'_{2}\one
\rmx
- \lambda_{max}\|D'_{1}\one + M_{11}^{-1}M_{12}D'_{2}\one
- \delta M_{11}^{-1}\one
\|\one
+
\lmx 
\delta \one\\
-\delta \|M_{12}^{T}M_{11}^{-1}\one\|\one
\rmx 
\\
&\geq 
\lmx 0 \\
SD'_{2}\one
\rmx
-
\frac{3}{4}\delta \one
+
\lmx 
\delta \one\\
-\delta \kappa^{1/2}n\one
\rmx 
\quad \text{(using Lemmas \ref{lem:aux1} and \ref{lem:aux2})}
\\
&\geq 
\lmx 0 \\
SD'_{2}\one
-2\delta \kappa^{1/2} n\one
\rmx \\
&=
\lmx 0 \\
S\Sigma^{-1/2}D_{R}\one
-\frac{1}{12}\lambda_{min}^{1/2}\one
\rmx \\
&\geq \lmx 0 \\
S\Sigma^{-1/2}D_{R}\one
-\frac{1}{12}S_{D}^{1/2}\one
\rmx
\quad \text{(using Fact \ref{fact:mmatrix}.\ref{fact:schurdiag})} 
\\
&\geq \lmx 0 \\
S\Sigma^{-1/2}D_{R}\one
-\frac{1}{6}S_{D}^{1/2}\tilde{S}_{D}^{1/2}\one
\rmx
\quad \text{(using Lemma \ref{lem:scaledschur})} 
\end{align*}

\end{proof}

\begin{lemma}\label{lem:aux2}
For all positive vectors $\vv$,
\[
\|M_{12}^{T}M_{11}^{-1}\vv\|
\leq 
\kappa^{1/2}n^{1/2}\norm{\vv}
\]
\end{lemma}

\begin{proof}
Define 
$c = \lambda_{min}^{-1}\kappa^{-1/2}n^{-1/2}\norm{\vv } = 
\lambda_{max}^{-1}\kappa^{1/2}n^{-1/2}\norm{\vv}$.
\begin{align*}
\|M_{12}^{T}M_{11}^{-1}\vv\|
\leq 
\|M_{12}^{T}M_{11}^{-1}\vv\|_{1}
&=
-\one^{T}M_{12}^{T}M_{11}^{-1}\vv 
\qquad \text{(by Fact \ref{fact:mmatrix}, $M_{11}^{-1}$ and $-M_{12}$
are nonnegtive)}
\\
&= 
\frac{1}{2c}\left(
\vv^{T}M_{11}^{-1}\vv
+ c^{2}\one^{T}M_{22}\one
- \lmx \vv^{T}M_{11}^{-1} & c\one^{T} \rmx 
M
\lmx M_{11}^{-1}\vv \\ c\one \rmx
\right) \\
&\leq  
\frac{1}{2c}\left(
\vv^{T}M_{11}^{-1}\vv
+ c^{2}\one^{T}M_{22}\one
\right) \\
&\leq  
\frac{1}{2}\left(
\frac{\norm{\vv}^{2}}{c\lambda_{min}} + c\lambda_{max}n
\right)
=
\kappa^{1/2}n^{1/2}\norm{\vv}
\end{align*}

\end{proof}


\section{Solving Matrices from the Interior-Point Method}\label{app:reduce}

In the interior-point algorithm, we need to solve
matrices of the form 
\[
M + \vv \vv^{T} = 
\lmx AD_{1}^{2}A^{T} + D_{2}^{2} & AD_{1}^{2} \\ 
D_{1}^{2}A^{T} & D_{1}^{2} + D_{3}^{2} \rmx
+ \vv\vv^{T}
\]
where $A$ is an $n \times m$ matrix with entries bounded by $U$ 
in absolute value,
$AA^{T}$ is an M-matrix, and $D_{1},D_{2},D_{3}$ 
are positive diagonal matrices. 
We show how to do this using
our $\mathtt{MMatrixSolve}$ algorithm.

Consider the Schur complement of $M$:
\[
M_{S} = (AD_{1}^{2}A^{T} + D_{2}^{2}) - 
AD_{1}^{2}(D_{1}^{2}+D_{3}^{2})^{-1}D_{1}^{2}A^{T} =
AD_{1}^{2}D_{3}^{2}(D_{1}^{2}+D_{3}^{2})^{-1}A^{T} + D_{2}^{2} =
A_{S}A_{S}^{T} 
\]
where 
$A_{S} = 
\lmx AD_{1}D_{3}(D_{1}^{2}+D_{3}^{2})^{-1/2} & D_{2} \rmx 
$.
Note that $M_{S}$ is also an M-matrix,
and that the eigenvalues of $M_{S}$ fall in the range
$[d_{min}^{2},d_{max}^{2}(U\sqrt{nm}+1)]$
where $d_{min}$ and $d_{max}$
are respectively the smallest and largest diagonal entry in $D_{1},D_{2},D_{3}$.

We can build an solver for systems in $M$ from 
a solver for systems in $M_{S}$,
by using the following easily verifiable
property of the Schur complement:
\begin{lemma}\label{lem:schurnorm}
For $M = \lmx M_{11} & M_{12} \\ M_{12}^{T} & M_{22} \rmx$
and Schur complement
$M_{S} = M_{11}-M_{12}M_{22}^{-1}M_{12}^{T}$, we have
\[
\norm{
\lmx \xx_{1} \\ \xx_{2} \rmx -
M^{-1}\lmx \bb_{1} \\ \bb_{2} \rmx 
}_{M} = 
\norm{\xx_{1} - 
M_{S}^{-1}
(\bb_{1}-M_{12}M_{22}^{-1}\bb_{2})}_{M_{S}}
+
\norm{\xx_{2} - 
M_{22}^{-1}(\bb_{2} - M_{12}^{T}\xx_{1}) }_{M_{22}}
\]
\end{lemma}

Then, to solve systems in $M+\vv \vv^{T}$, 
we can use the Sherman-Morrison formula:
\[
(M+\vv \vv^{T})^{-1} = M^{-1} -
\frac{M^{-1}\vv \vv^{T}M^{-1}}{1+\vv^{T}M^{-1}\vv}
\]

In particular, we give the following algorithm, which runs in time
$\softO{m\log \frac{\kappa \norm{\vv}}{\epsilon}}$:

\mybox{
$\xx = 
\mathtt{Solve}(M 
+\vv\vv^{T},\bb,\epsilon)$
\qquad 
where
$M = 
\lmx AD_{1}^{2}A^{T} + D_{2}^{2} & AD_{1}^{2} \\ 
D_{1}^{2}A^{T} & D_{1}^{2} + D_{3}^{2} \rmx
$
and
$\bb = \lmx \bb_{1} \\ \bb_{2} \rmx $
and
$\vv = \lmx \vv_{1} \\ \vv_{2} \rmx $
\begin{itemize}
\item Define
$\epsilon_{1} = \frac{\epsilon}{2}(1+\vv^{T}M^{-1}\vv)^{-1}$
and 
$\epsilon_{2} = \min\setof{\frac{1}{2},\frac{\epsilon}{14}(1+\vv^{T}M^{-1}\vv)^{-1}}$
\item $\yy' = \mathtt{MMatrixSolve}
\left(A_{S},
\bb_{1}-AD_{1}^{2}(D_{1}^{2}+D_{3}^{2})^{-1}\bb_{2},\epsilon_{1},
d_{min}^{2},d_{max}^{2}(U\sqrt{nm}+1)\right)$
\item 
$\yy = \lmx \yy' \\ 
(D_{1}^{2}+D_{3}^{2})^{-1}(\bb_{2} - D_{1}^{2}A^{T}\yy') \rmx $
\item $\zz' = \mathtt{MMatrixSolve}
\left(A_{S},
\vv_{1}-AD_{1}^{2}(D_{1}^{2}+D_{3}^{2})^{-1}\vv_{2},\epsilon_{2},
d_{min}^{2},d_{max}^{2}(U\sqrt{nm}+1)\right)$
\item 
$\zz = \lmx \zz' \\ 
(D_{1}^{2}+D_{3}^{2})^{-1}(\vv_{2} - D_{1}^{2}A^{T}\zz') \rmx $
\item Return $\xx = \yy - \frac{\zz\zz^{T}\bb}{1+\vv^{T}\zz}$
\end{itemize}
}

\begin{lemma}
$\xx = 
\mathtt{Solve}(M 
+\vv\vv^{T},\bb,\epsilon)$ satisfies
\[
\norm{\xx - (M+\vv \vv^{T})^{-1}\bb}_{M+\vv \vv^{T}} < 
\epsilon\norm{(M+\vv \vv^{T})^{-1}\bb}_{M+\vv \vv^{T}}
\]
\end{lemma}
\begin{proof}
We first show that 
$\norm{\yy - M^{-1}\bb}_{M} \leq \epsilon_{1}\norm{M^{-1}\bb}_{M}$:
\begin{align*}
\norm{\yy - M^{-1}\bb}_{M}
&=
\norm{\yy' - M_{S}^{-1}(\bb_{1}-AD_{1}^{2}(D_{1}^{2}+D_{3}^{2})^{-1}\bb_{2})}_{M_{S}} 
\quad \text{(by Lemma \ref{lem:schurnorm})}\\
&\leq 
\epsilon_{1}\norm{M_{S}^{-1}(\bb_{1}-AD_{1}^{2}(D_{1}^{2}+D_{3}^{2})^{-1}\bb_{2})}_{M_{S}} 
\quad \text{(guaranteed by $\mathtt{MMatrixSolve}$)}\\
&=
\epsilon_{1}\left(\norm{M^{-1}\bb}_{M} -
\norm{M_{22}^{-1}\bb_{2}}_{M_{22}} \right)
\quad \text{(by Lemma \ref{lem:schurnorm})}
\\
&\leq 
\epsilon_{1}\norm{M^{-1}\bb}_{M}\\
\end{align*}
By the same reasoning, 
$\norm{\zz - M^{-1}\vv}_{M} \leq \epsilon_{2}\norm{M^{-1}\vv}_{M}$.

Next, let us define the inner product
$\inprod{\vv_{1}}{\vv_{2}}_{M} = \vv_{1}^{T}M\vv_{2}$.
We will use repeatedly the
inequality 
$\sizeof{\inprod{\vv_{1}}{\vv_{2}}_{M}} \leq \norm{\vv_{1}}_{M}\norm{\vv_{2}}_{M}$

Recall that we return the value 
$\xx = \yy - \frac{\zz \zz^{T}\bb}{1+\vv^{T}\zz}$.
So we begin by analyzing the expressions $\zz \zz^{T}\bb$ and $\vv^{T}\zz$:
\begin{align}
\norm{\zz \zz^{T}\bb - M^{-1}\vv\vv^{T}M^{-1}\bb}_{M}
&\leq 
\norm{\zz \zz^{T}\bb - \zz\vv^{T}M^{-1}\bb}_{M} + 
\norm{\zz \vv^{T}M^{-1}\bb - M^{-1}\vv\vv^{T}M^{-1}\bb}_{M} 
\notag \\
&=
\sizeof{\inprod{\zz - M^{-1}\vv}{M^{-1}\bb}_{M}}\norm{\zz}_{M} + 
\sizeof{\inprod{M^{-1}\vv}{M^{-1}\bb}_{M}}\norm{\zz - M^{-1}\vv}_{M} 
\notag \\
&\leq 
\norm{\zz - M^{-1}\vv}_{M}\norm{M^{-1}\bb}_{M}\norm{\zz}_{M} + 
\norm{M^{-1}\vv}_{M}\norm{M^{-1}\bb}_{M}\norm{\zz - M^{-1}\vv}_{M}
\notag \\
&=
\norm{\zz - M^{-1}\vv}_{M}\norm{M^{-1}\bb}_{M}
\left(\norm{\zz}_{M} + \norm{M^{-1}\vv}_{M} \right)
\notag \\
&\leq 
\norm{\zz - M^{-1}\vv}_{M}\norm{M^{-1}\bb}_{M}
\left(\norm{\zz-M^{-1}\vv}_{M} + 2\norm{M^{-1}\vv}_{M} \right)
\notag \\
&\leq \epsilon_{2}(\epsilon_{2}+2)
\norm{M^{-1}\vv}_{M}^{2}\norm{M^{-1}\bb}_{M}
\label{eq:zzb}
\end{align}
\begin{align}
|\vv^{T}\zz - \vv^{T}M^{-1}\vv|
&=
\sizeof{\inprod{M^{-1}\vv}{\zz - M^{-1}\vv}_{M}}
\notag \\
&\leq \norm{M^{-1}\vv}_{M}\norm{\zz - M^{-1}\vv}_{M}
\notag \\
&\leq \epsilon_{2}\norm{M^{-1}\vv}_{M}^{2}
\notag \\
&= \epsilon_{2}(\vv^{T}M^{-1}\vv)
\label{eq:vz}
\end{align}

We thus have:

\begin{align}
&\norm{\xx - (M+\vv \vv^{T})^{-1}\bb}_{M} \\
&=
\norm{\left( \yy - \frac{\zz \zz^{T}\bb}{1+\vv^{T}\zz} \right)-
\left(M^{-1}\bb - \frac{M^{-1}\vv \vv^{T}M^{-1}\bb}{1+\vv^{T}M^{-1}\vv}\right)}_{M} 
\notag \\
&\leq 
\norm{\yy - M^{-1}\bb}_{M} +
\norm{\frac{\zz \zz^{T}\bb}{1+\vv^{T}\zz} -
\frac{M^{-1}\vv \vv^{T}M^{-1}\bb}{1+\vv^{T}\zz}}_{M} +
\norm{\frac{M^{-1}\vv \vv^{T}M^{-1}\bb}{1+\vv^{T}\zz} -
\frac{M^{-1}\vv \vv^{T}M^{-1}\bb}{1+\vv^{T}M^{-1}\vv}}_{M} 
\notag \\
&= 
\norm{\yy - M^{-1}\bb}_{M} +
\frac{1}{1+\vv^{T}\zz}
\left(
\norm{\zz \zz^{T}\bb -
M^{-1}\vv \vv^{T}M^{-1}\bb}_{M} +
\frac{|\vv^{T}\zz -\vv^{T}M^{-1}\vv|}{1+\vv^{T}M^{-1}\vv}
\norm{M^{-1}\vv \vv^{T}M^{-1}\bb}_{M}
\right)
\notag \\
&\leq  
\norm{\yy - M^{-1}\bb}_{M} +
\frac{1}{\vv^{T}\zz}
\left(
\norm{\zz \zz^{T}\bb -
M^{-1}\vv \vv^{T}M^{-1}\bb}_{M} +
\frac{|\vv^{T}\zz -\vv^{T}M^{-1}\vv|}{\vv^{T}M^{-1}\vv}
\norm{M^{-1}\vv \vv^{T}M^{-1}\bb}_{M}
\right)
\notag \\
&\leq 
\norm{\yy - M^{-1}\bb}_{M} +
\frac{1}{(1-\epsilon_{2})\vv^{T}M^{-1}\vv}
\left(
\epsilon_{2}(\epsilon_{2}+2)\norm{M^{-1}\vv}_{M}^{2}\norm{M^{-1}\bb}_{M} + 
\epsilon_{2}
\norm{M^{-1}\vv \vv^{T}M^{-1}\bb}_{M}
\right)
\notag \\
&\qquad \text{(by equations \ref{eq:zzb} and \ref{eq:vz})}
\notag \\
&=
\norm{\yy - M^{-1}\bb}_{M} +
\frac{
\epsilon_{2}(\epsilon_{2}+2)\norm{M^{-1}\vv}_{M}^{2}\norm{M^{-1}\bb}_{M} + 
\epsilon_{2}
\sizeof{\inprod{M^{-1}\vv}{M^{-1}\bb}_{M}}\norm{M^{-1}\vv}_{M}
}{(1-\epsilon_{2})\norm{M^{-1}\vv}_{M}^{2}}
\notag \\
&\leq 
\norm{\yy - M^{-1}\bb}_{M} +
\frac{
\epsilon_{2}(\epsilon_{2}+2)\norm{M^{-1}\vv}_{M}^{2}\norm{M^{-1}\bb}_{M} + 
\epsilon_{2}
\norm{M^{-1}\vv}_{M}^{2}\norm{M^{-1}\bb}_{M}
}{(1-\epsilon_{2})\norm{M^{-1}\vv}_{M}^{2}}
\notag \\
&=
\norm{\yy - M^{-1}\bb}_{M} +
\frac{\epsilon_{2} (\epsilon_{2}+3)}{1-\epsilon_{2}}\norm{M^{-1}\bb}_{M}
\notag \\
&\leq 
\left(
\epsilon_{1} + 
\frac{\epsilon_{2} (\epsilon_{2}+3)}{1-\epsilon_{2}}
\right)
\norm{M^{-1}\bb}_{M}
\notag \\
&\leq 
\epsilon (1+\vv^{T}M^{-1}\vv )^{-1}
\norm{M^{-1}\bb}_{M}
\label{eq:almostdone}
\end{align}

So we conclude
\begin{align*}
\norm{\xx -(M+\vv \vv^{T})^{-1}\bb}_{M+\vv \vv^{T}}
&\leq 
(1+\vv^{T}M^{-1}\vv)^{1/2}\norm{\xx -(M+\vv \vv^{T})^{-1}\bb}_{M}
\qquad \text{by Lemma \ref{lem:rankoneupdate}(i)}
\\
&\leq 
\epsilon (1+\vv^{T}M^{-1}\vv )^{-1/2}
\norm{M^{-1}\bb}_{M}
\qquad \text{by equation (\ref{eq:almostdone})}
\\
&=
\epsilon (1+\vv^{T}M^{-1}\vv )^{-1/2}
\norm{\bb}_{M^{-1}}
\\
&\leq 
\epsilon \norm{\bb}_{(M+\vv \vv^{T})^{-1}}
\qquad \text{by Lemma \ref{lem:rankoneupdate}(ii)}
\\
&=
\epsilon \norm{(M+\vv \vv^{T})^{-1}\bb}_{M+\vv \vv^{T}}
\end{align*}
\end{proof}

\begin{lemma}\label{lem:rankoneupdate}
 For all vectors $\vv$, $\ww$, and symmetric positive definite $M$:
\begin{align*}
&\text{(i)}&
\norm{\ww}_{M+\vv \vv^{T}}
&\leq 
\norm{\ww}_{M}(1+\vv^{T}M^{-1}\vv)^{1/2}
\\
&\text{(ii)}&
\norm{\ww}_{(M+\vv \vv^{T})^{-1}}
&\geq 
\norm{\ww}_{M^{-1}}(1+\vv^{T}M^{-1}\vv)^{-1/2}
\end{align*}
\end{lemma}

\begin{proof}[Proof of (i)]
\begin{align*}
\norm{\ww}_{M+\vv \vv^{T}}
&=
(\ww^{T}(M+\vv \vv^{T})\ww)^{1/2}\\
&=
(\ww^{T}M\ww + (\ww^{T}\vv)^{2})^{1/2}\\
&=
(\norm{\ww}_{M}^{2} + \inprod{\ww}{M^{-1}\vv}_{M}^{2})^{1/2}\\
&\leq 
(\norm{\ww}_{M}^{2} + \norm{\ww}_{M}^{2}\norm{M^{-1}\vv}_{M}^{2})^{1/2}\\
&=
\norm{\ww}_{M}(1+\norm{M^{-1}\vv}_{M}^{2})^{1/2}
\end{align*}
\end{proof}
\begin{proof}[Proof of (ii)]
\begin{align*}
\norm{\ww}_{(M+\vv \vv^{T})^{-1}}
&=
(\ww^{T}(M+\vv \vv^{T})^{-1}\ww )^{1/2}\\
&=
\left(\ww^{T}
\left(M^{-1} - \frac{M^{-1}\vv \vv^{T}M^{-1}}{1+\vv^{T}M^{-1}\vv}\right)
\ww\right)^{1/2}\\
&=
\left(\ww^{T}M^{-1}\ww - \frac{\ww^{T}M^{-1}\vv \vv^{T}M^{-1}\ww}{1+\vv^{T}M^{-1}\vv} \right)^{1/2}\\
&=
\left(\norm{\ww}_{M^{-1}}^{2} - \frac{\inprod{\ww}{\vv}_{M^{-1}}^{2}}{1+\norm{\vv}_{M^{-1}}^{2}} \right)^{1/2}\\
&\geq 
\left(\norm{\ww}_{M^{-1}}^{2} - \frac{\norm{\ww}_{M^{-1}}^{2}\norm{\vv}_{M^{-1}}^{2}}{1+\norm{\vv}_{M^{-1}}^{2}} \right)^{1/2}\\
&=
\norm{\ww}_{M^{-1}}\left(1 - \frac{\norm{\vv}_{M^{-1}}^{2}}{1+\norm{\vv}_{M^{-1}}^{2}} \right)^{1/2}\\
&=
\norm{\ww}_{M^{-1}}\left(1+\norm{\vv}_{M^{-1}}^{2}\right)^{-1/2}
\end{align*}
\end{proof}


\section{Interior-Point Method using an Approximate Solver}\label{app:intpt}

Throughout this section,
we take $\mathtt{Solve}$ to be an algorithm such that
$\xx = \mathtt{Solve}(M,\bb,\epsilon)$
satisfies
\[
\norm{\xx - M^{-1}\bb}_{M} \leq 
\epsilon \norm{M^{-1}\bb}
\]

We use the notational convention that
$S$ denotes the diagonal matrix whose diagonal is $\ss$.
The same applies for $X$ and $\xx$, etc.

$\one_{k}$ denotes the all-ones vector of length $k$.

$\interior{\Omega}$ denotes the interior of polytope $\Omega$.

We are given a canonical primal linear program
\[
\zz^{*} = \min_{\xx}\setof{\cc^{T}\xx : A\xx = \bb ; \xx \geq 0}
\]
which has the same solution as the dual linear program
\[
\zz^{*} = \max_{(\yy , \ss)}\setof{\bb^{T}\yy : A^{T}\yy + \ss = \cc; \ss \geq 0}
\]
where $A$ is an $n\times m$ matrix,
$\xx,\ss,\cc$ are length $m$, and $\yy,\bb$ are length $n$, and $m \geq n$.
(Unfortunately, this use of $n$ and $m$ is reversed from
the standard linear programming convention.  We do this
to be consistent with the standard graph-theory convention
that we use throughout the paper.)

We let $\Omega^{D}$ denote the dual polytope 
\[
\Omega^{D} = \setof{(\yy ,\ss) : A^{T}\yy + \ss = \cc; \ss \geq 0}
\]
so we can write the solution to the linear program as 
$\zz^{*} = \max_{(\yy ,\ss)\in \Omega^{D}} \bb^{T}\yy$.

In this appendix, we present an 
$\mathtt{InteriorPoint}$ algorithm based on that of Renegar \cite{Renegar},
modified to use an approximate solver.
Our analysis follows that found in \cite{YeBook}.

\newtheorem*{theorem:intptalg}{Theorem \ref{thm:intptalg}}
\begin{theorem:intptalg}
$\xx = \mathtt{InteriorPoint}(A,\bb,\cc,\lambda_{min},T,\yy^{0},\epsilon)$ takes input
that satisfy
\begin{itemize}
\item $A$ is an $n\times m$ matrix; 
$\bb$ is a length $n$ vector;
$\cc$ is a length $m$ vector
\item $AA^{T}$ is positive definite,
and $\lambda_{min} > 0$ is a lower bound on the eigenvalues of $AA^{T}$
\item $T>0$ is an upper bound on the 
absolute values of the dual coordinates, i.e.

\qquad
$\norm{\yy}_{\infty} < T$ and $\norm{\ss}_{\infty} < T$
for all $(\yy,\ss)$ that satisfy $\ss = \cc - A^{T}\yy \geq 0$

\item initial point $\yy^{0}$ is a length $n$ vector where
$A^{T}\yy^{0} < \cc$
\item error parameter $\epsilon$ satisfies $0 < \epsilon < 1$
\end{itemize}
and returns $\xx > 0$ satisfying
$\norm{A\xx - \bb} \leq \epsilon$ and $z^{*} < \cc^{T}\xx < z^{*} + \epsilon$.

Let us define
\begin{itemize}
\item $U$ is the largest absolute value of any entry in $A,\bb ,\cc$
\item $s^{0}_{min}$ is the smallest entry of $\ss^{0} = \cc - A^{T}\yy^{0}$
\end{itemize}
Then the algorithm 
makes $\bigO{\sqrt{m}\log \frac{TUm}{\lambda_{min}s^{0}_{min}\epsilon}}$
calls to the approximate solver, of the form
\[
\mathtt{Solve}\left(AS^{-2}A^{T} + \vv\vv^{T},\cdot,\epsilon'\right)
\]
where $S$ is a positive diagonal matrix with condition number
$\bigO{\frac{T^{2}Um^{2}}{\epsilon}}$, and $\vv,\epsilon'$ satisfy
\[
\log \frac{\norm{\vv}}{\epsilon'}
= 
\bigO{\log \frac{TUm}{s^{0}_{min}\epsilon}}
\]
\end{theorem:intptalg}

\subsection{The Analytic Center}

Standard interior-point methods focus on a particular point
in the interior of the dual polytope.
This point, called 
the {\em analytic center},
is the point that
maximizes the product of the slacks, 
i.e. the product of the elements of $\ss$.
For the purpose of our analysis, we use the following
equivalent definition of the analytic center:
\begin{fact}[see {\cite[\S 3.1]{YeBook}}]
The {\bf analytic center} of 
$\Omega^{D} = \setof{(\yy ,\ss) : A^{T}\yy + \ss = \cc; \ss \geq 0}$
is the 
unique point $(\overstar{\yy},\overstar{\ss})\in \interior{\Omega}^{D}$ that 
satisfies $\eta_{A}(\overstar{\ss}) = 0$, where we define
\begin{align*}
\xx_{A}(\ss) 
&= 
S^{-1}(I-S^{-1}A^{T}(AS^{-2}A^{T})^{-1}AS^{-1})\one_{m}
\\
\eta_{A}(\ss) 
&=
\norm{S\xx_{A}(\ss) - \one_{m}} 
=
\norm{S^{-1}A^{T}(AS^{-2}A^{T})^{-1}AS^{-1}\one_{m}}
=
\norm{AS^{-1}\one_{m}}_{(AS^{-2}A^{T})^{-1}}
\end{align*}
\end{fact}
These definitions of $\xx_{A}$ and $\eta_{A}$ satisfying the following
properties:
\begin{lemma}\label{lem:acprops}
Let $(\overstar{\yy},\overstar{\ss})$ be the
analytic center of $\Omega^{D}$.
For any point $(\yy ,\ss ) \in \interior{\Omega}^{D}$ we have
\begin{enumerate}[label=(\roman{*}),ref=\ref{lem:acprops}(\roman{*})]
\item \label{lem:axeqzero}
$A\xx_{A}(\ss ) = 0$
\item \label{lem:xpos}
$\eta_{A}(\ss) < 1$ implies $\xx_{A}(\ss) > 0$
\item \label{lem:xinvs}
$\xx_{A}(\overstar{\ss}) = \overstar{S}^{-1}\one_{m}$
\item \label{lem:etamin}
For all $\xx$ satisfying $A\xx = 0$, it holds that
$\norm{S\xx - \one_{m}} \geq \eta_{A}(\ss)$
\end{enumerate}
\end{lemma}
The first three properties are straightforward from the definition.  
We present a proof of the last:
\begin{proof}[Proof of \ref{lem:etamin}]

Note that $S\xx_{A}(\ss) - \one_{m}$
is orthogonal to
$S(\xx-\xx_{A}(\ss))$, because
\begin{align*}
\inprod{S\xx_{A}(\ss) - \one_{m}}{S(\xx-\xx_{A}(\ss))}
&=
\inprod{S^{-1}A^{T}(AS^{-2}A^{T})^{-1}AS^{-1}\one_{m}}
{S(\xx-\xx_{A}(\ss))}
\\
&=
\inprod{(AS^{-2}A^{T})^{-1}AS^{-1}\one_{m}}
{A(\xx-\xx_{A}(\ss))}
\\
&=
\inprod{(AS^{-2}A^{T})^{-1}AS^{-1}\one_{m}}
{0} 
\\
&= 
0
\end{align*}
We thus have
\[
\norm{S\xx - \one_{m}}
=
\norm{S\xx_{A}(\ss) - \one_{m} + S(\xx-\xx_{A}(\ss))}
\geq 
\norm{S\xx_{A}(\ss) - \one_{m}}
= 
\eta_{A}(\ss)
\]

\end{proof}

It will be useful to note that
the slacks of the analytic center cannot be too small.
We can bound the slacks of the analytic center 
away from zero as follows:
\begin{lemma}[compare {\cite[Thm 2.6]{YeBook}}]\label{lem:boundedslacks}
Let 
$(\overstar{\yy},\overstar{\ss})$ be the analytic center of
$\Omega^{D}$.
For every $(\yy , \ss )\in \Omega^{D}$,
we have $\overstar{\ss} > \frac{1}{m}\ss$
\end{lemma}
\begin{proof}
\begin{align*}
\norm{\overstar{S}^{-1}\ss}_{\infty}
\leq 
\one_{m}^{T}\overstar{S}^{-1}\ss 
&=
\one_{m}^{T}\overstar{S}^{-1}\overstar{\ss} +
\one_{m}^{T}\overstar{S}^{-1}(\ss - \overstar{\ss}) 
\\
&=
m+
\one_{m}^{T}\overstar{S}^{-1}(\ss - \overstar{\ss}) 
\\
&=
m+
\one_{m}^{T}\overstar{S}^{-1}
\left((\cc - A^{T}\yy) - (\cc - A^{T}\overstar{\yy}) \right)
\\
&=
m+
\one_{m}^{T}\overstar{S}^{-1}A^{T}(\overstar{\yy} - \yy)
\\
&=
m
\end{align*}
where we know from Lemmas \ref{lem:axeqzero} and \ref{lem:xinvs}
that $A\overstar{S}^{-1}\one_{m} = 0$
\end{proof}

Let us note that
a point $(\yy ,\ss )\in \interior{\Omega}^{D}$ that satisfies
$\eta_{A}(\ss) < 1$ is close to the analytic center,
in the sense that the slacks $\ss$
are bounded by a constant ratio from the slacks
of the analytic center:
\begin{lemma}[{\cite[Thm 3.2(iv)]{YeBook}}]\label{lem:nearcenter}
Suppose $(\yy , \ss )\in \interior{\Omega}^{D}$ 
satisfies $\eta_{A}(\ss) = \eta < 1$
and let 
$(\overstar{\yy},\overstar{\ss})$ be the analytic center of
$\Omega^{D}$.
Then 
$\norm{S^{-1}\overstar{\ss} - \one_{m}} \leq  \frac{\eta}{1-\eta}$.
\end{lemma}

If $(\yy ,\ss )\in \interior{\Omega}^{D}$
is sufficiently close to the analytic center
(as measured by $\eta_{A}$),
then with a single call to the approximate solver, we can 
take a Newton-type step to find
a point even closer to the analytic center.
This $\mathtt{NewtonStep}$ procedure is 
presented in Figure \ref{fig:newtonstep}.
\begin{figure}[t]
\mybox{
$\yy^{+} = \mathtt{NewtonStep}(A,\cc,\yy)$
\begin{itemize}
\item Let $\ss = \cc - A^{T}\yy$
\item Let 
$\dd_{y} = \mathtt{Solve}(AS^{-2}A^{T},-AS^{-1}\one_{m},\epsilon_{3})$
where $\epsilon_{3} = \frac{1}{20(\sqrt{m} + 1)}$
\item Return $\yy^{+} = \yy + (1-\epsilon_{3} )\dd_{y}$
\end{itemize}
}
\caption{Procedure for stepping closer to the analytic center}
\label{fig:newtonstep}
\end{figure}

In the first part of the following lemma,
we prove that the point returned by $\mathtt{NewtonStep}$
is indeed still inside the dual polytope.
In the second part, 
we show how close the new point is to the analytic center:
\begin{lemma}[compare {\cite[Thm 3.3]{YeBook}}]\label{lem:newtonstep}
Suppose $(\yy,\ss )\in \interior{\Omega}^{D}$ 
satisfies $\eta_{A} (\ss ) = \eta < 1$.

Let $\yy^{+} = \mathtt{NewtonStep}(A,\cc,\yy)$
and $\ss^{+} = \cc - A^{T}\yy^{+}$

Then (i) $\ss^{+} > 0$
and (ii) $\eta_{A} (\ss^{+}) \leq \eta^{2} + \frac{1}{20}\eta$
\end{lemma}
\begin{proof}
(i) 
The solver guarantees that
\[
\norm{\dd_{y} + (AS^{-2}A^{T})^{-1}AS^{-1}\one}_{AS^{-2}A^{T}}
\leq \epsilon_{3}
\norm{(AS^{-2}A^{T})^{-1}AS^{-1}\one}_{AS^{-2}A^{T}}
= \epsilon_{3}\cdot 
\eta
\]
or equivalently
\begin{equation}\label{eq:newtonsolve}
\norm{S^{-1}A^{T}\dd_{y} + S^{-1}A^{T}(AS^{-2}A^{T})^{-1}AS^{-1}\one}
\leq \epsilon_{3}
\norm{S^{-1}A^{T}(AS^{-2}A^{T})^{-1}AS^{-1}\one}
= \epsilon_{3}\cdot 
\eta
\end{equation}
and so
\[
\norm{S^{-1}A^{T}\dd_{y}}
\leq (1+\epsilon_{3})
\norm{S^{-1}A^{T}(AS^{-2}A^{T})^{-1}AS^{-1}\one}
= (1+\epsilon_{3})\eta < 1 + \epsilon_{3}
\]

We thus have
\begin{align*}
\norm{S^{-1}\ss^{+} - \one}
&=
\norm{S^{-1}\left(\ss - (1-\epsilon_{3})A^{T}\dd_{y} \right) - \one} \\
&=
(1-\epsilon_{3})\norm{S^{-1}A^{T}\dd_{y}} \\
&\leq 
(1-\epsilon_{3})(1+\epsilon_{3}) < 1
\end{align*}
Thus $S^{-1}\ss^{+}$ is positive and so is $\ss^{+}$.

\medskip\noindent
(ii)
Let $\xx = \xx_{A} (\ss )$ and $\xx^{+} = \xx_{A} (\ss^{+})$.
We have
\begin{align*}
\eta_{A}(\ss^{+})
&\leq 
\norm{X\ss^{+} - \one_{m}}
\qquad \text{(by Lemma \ref{lem:etamin})}
\\
&=
\norm{X(\ss - (1-\epsilon_{3})A^{T}\dd_{y}) - \one_{m}} \\
&= 
\norm{(1-\epsilon_{3})XS(X\ss - \one_{m} - S^{-1}A^{T}\dd_{y})
- (1-\epsilon_{3})(XS-I)(X\ss - \one_{m}) 
+ \epsilon_{3}(X\ss -\one_{m})
}
\\
&\leq 
(1-\epsilon_{3})\norm{XS(S^{-1}A^{T}\dd_{y} - S\xx + \one_{m})} +
(1-\epsilon_{3})\norm{(XS-I)(X\ss -\one_{m})} + 
\epsilon_{3}\norm{(X\ss-\one_{m})}
\\
&\leq 
(1-\epsilon_{3})\norm{S\xx}
\norm{S^{-1}A^{T}\dd_{y} - S\xx + \one_{m}} +
(1-\epsilon_{3})\norm{S\xx -\one_{m}}^{2} + 
\epsilon_{3}\norm{S\xx-\one_{m}}
\\
&\qquad 
\text{(using the relation 
$\norm{V\ww} \leq 
\norm{\vv}_{\infty}\norm{\ww} \leq 
\norm{\vv}\norm{\ww}$)
}
\\
&\leq 
(1-\epsilon_{3})(\norm{S\xx-\one_{m}}+\norm{\one_{m}})\norm{S^{-1}A^{T}\dd_{y} - S\xx + \one_{m}}
+ (1-\epsilon_{3})\norm{S\xx - \one_{m}}^{2} + \epsilon_{3}\norm{S\xx - \one_{m}}
\\
&=
(1-\epsilon_{3})(\eta+\sqrt{m})\norm{S^{-1}A^{T}\dd_{y} - S\xx + \one_{m}}
+ (1-\epsilon_{3})\eta^{2} + \epsilon_{3}\eta
\\
&=
(1-\epsilon_{3})(\eta+\sqrt{m})\norm{S^{-1}A^{T}\dd_{y} + S^{-1}A^{T}(AS^{-2}A^{T})^{-1}AS^{-1}\one_{m}}
+ (1-\epsilon_{3})\eta^{2} + \epsilon_{3}\eta
\\
&\leq 
(1-\epsilon_{3})(\eta+\sqrt{m})\epsilon_{3}\eta
+ (1-\epsilon_{3})\eta^{2} + \epsilon_{3}\eta
\qquad \text{(by equation \ref{eq:newtonsolve})}
\\
&\leq 
\epsilon_{3}(\eta+\sqrt{m})\eta
+ (1-\epsilon_{3})\eta^{2} + \epsilon_{3}\eta
\\
&= 
\eta^{2} + \epsilon_{3}(\sqrt{m}+1)\eta
\\
&= 
\eta^{2} + \frac{1}{20}\eta
\end{align*}
\end{proof}

\subsection{The Path-Following Algorithm}

In a path-following algorithm,
we modify the dual polytope 
$\Omega^{D} = \setof{(\yy,\ss):A^{T}\yy + \ss = \cc ; \ss \geq 0}$
by adding an additional contraint
$\bb^{T}\yy \geq z$, where $z \leq z^{*}$.
As we let $z$ approach $z^{*}$,
the center of the polytope approaches the solution to the 
dual linear program.

Letting $s_{gap} = \bb^{T}\yy - z$ denote the new slack variable,
we define the modified polytope:
\[
\Omega^{D}_{\bb,z} =
\setof{\left(\yy , \ss, s_{gap} \right): 
\lmx A^{T}\yy + \ss \\ -\bb^{T}\yy + s_{gap} \rmx 
= \lmx \cc \\ -z \rmx ; 
\ss,s_{gap} \geq 0}
\]
Using a trick of Renegar,
when we define the analytic center of $\Omega^{D}_{\bb,z}$,
we consider there to be $m$ copies of the slack $s_{gap}$, as follows:

\begin{definition}
The {\bf analytic center} of $\Omega^{D}_{\bb,z}$
is the
point $(\overstar{\yy},\overstar{\ss},\overstar{s}_{gap})\in \interior{\Omega}^{D}_{\bb,z}$, 
that 
satisfies $\tilde{\eta}(\overstar{\ss} ,\overstar{s}_{gap}) = 0$, where we define
\[
\tilde{\eta}(\ss,s_{gap}) = \eta_{\tilde{A}}(\tilde{\ss})
\qquad \text{where $\tilde{A} = \lmx A & -\bb \one_{m}^{T} \rmx$
and $\tilde{\ss} = \lmx \ss \\ s_{gap}\one_{m} \rmx $}
\]

The {\bf central path} is the set of analytic centers of the polytopes
$\setof{\Omega^{D}_{\bb,z}}_{z\leq z^{*}}$
\end{definition}

A path-following algorithm steps through a sequence of points near the 
central path, as $z$ increases towards $z^{*}$.
It is useful to note that
given any point on the central path,
we may easily construct a feasible primal solution $\xx$, as follows:
\begin{lemma}\label{lem:constructx}
Let $(\overstar{\yy},\overstar{\ss},\overstar{s}_{gap})$
be the analytic center of $\Omega^{D}_{\bb,z}$.
Then the vector $\xx = \frac{\overstar{s}_{gap}}{m}\overstar{S}^{-1}\one_{m}$
satisfies $A\xx = \bb $.
More generally,
for any
$(\yy,\ss,s_{gap})\in \interior{\Omega}^{D}_{\bb,z}$, 
the vector $\xx = \frac{s_{gap}}{m}S^{-1}\one_{m}$
satisfies 
\[
\norm{A\xx - \bb}_{(\tilde{A}\tilde{S}^{-2}\tilde{A}^{T})^{-1}}=
\frac{s_{gap}}{m}\cdot 
\tilde{\eta}(\ss,s_{gap})
\]
\end{lemma}
\begin{proof}
We prove the second assertion:
\begin{align*}
\norm{A\xx - \bb}_{(\tilde{A}\tilde{S}^{-2}\tilde{A}^{T})^{-1}}
&=
\frac{s_{gap}}{m}
\norm{
AS^{-1}\one_{m} 
- 
ms_{gap}^{-1}\bb
}_{(\tilde{A}\tilde{S}^{-2}\tilde{A}^{T})^{-1}}
\\
&=
\frac{s_{gap}}{m}
\norm{
\tilde{A}\tilde{S}^{-1}\one_{2m}
}_{(\tilde{A}\tilde{S}^{-2}\tilde{A}^{T})^{-1}}
\\
&=
\frac{s_{gap}}{m}\cdot 
\tilde{\eta}(\ss,s_{gap})
\end{align*}
The first assertion now follows from the definition of analytic center.
\end{proof}

\begin{figure}[t]
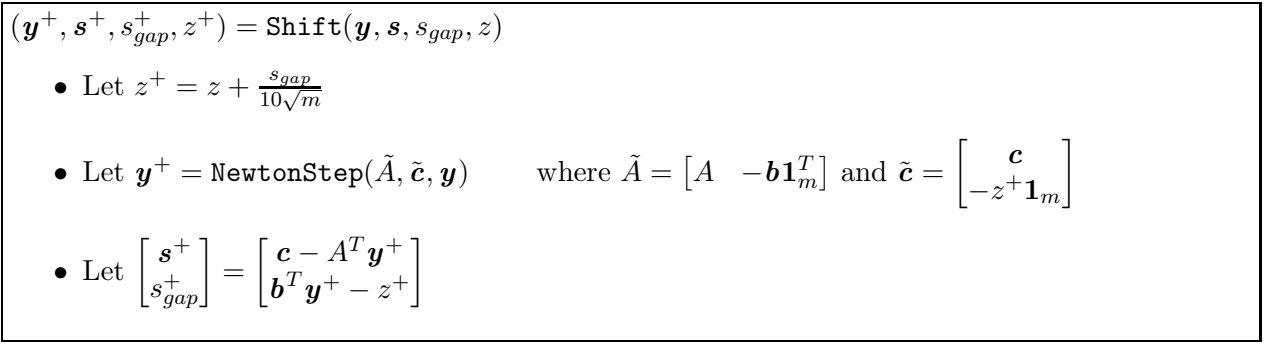

\mybox{
$(\yy^{+},\ss^{+},s_{gap}^{+},z^{+}) = 
\mathtt{Shift}(\yy,\ss,s_{gap},z)$
\begin{itemize}
\item Let $z^{+} = z + \frac{s_{gap}}{10\sqrt{m}}$
\item Let $\yy^{+} = 
\mathtt{NewtonStep}(\tilde{A},\tilde{\cc},\yy)$
\qquad 
where $\tilde{A} = \lmx A & -\bb \one_{m}^{T} \rmx $ 
and $\tilde{\cc} = \lmx \cc \\ -z^{+}\one_{m} \rmx$
\item Let
$\lmx \ss^{+} \\ s_{gap}^{+} \rmx =
\lmx \cc - A^{T}\yy^{+} \\ \bb^{T}\yy^{+} - z^{+} \rmx
$
\end{itemize}
}
\caption{Procedure for taking a step along the central path}
\label{fig:shift}
\end{figure}

Let us now describe how to take steps along the central path
using our approximate solver.  In Figure \ref{fig:shift}, 
we present the procedure
$\mathtt{Shift}$, which takes as input a value $z < z^{*}$ and
a point $(\yy,\ss,s_{gap})\in \interior{\Omega}^{D}_{\bb,z}$
satisfying $\eta(\ss,s_{gap}) \leq \frac{1}{10}$. 
The output is a new value $z^{+}$ that is closer to $z^{*}$,
and a new point $(\yy^{+},\ss^{+},s_{gap}^{+})\in \interior{\Omega}^{D}_{\bb,z^{+}}$
satisfying $\eta(\ss^{+},s_{gap}^{+}) \leq \frac{1}{10}$.
The procedure requires a single call to the solver.

Let us examine this procedure more closely.
After defining the incremented value $z^{+}$,
if we let $s_{gap}' = \bb^{T}\yy - z^{+} = s_{gap} - (z^{+} - z)$,
then $(\yy,\ss,s_{gap}')$
is a point in the shifted polytope
$\Omega^{D}_{\bb,z^{+}}$.
However this point
may be slightly farther away from the central path.
One call to the $\mathtt{NewtonStep}$ procedure
suffices to obtain a new point
$(\yy^{+},\ss^{+},s_{gap}^{+})\in \interior{\Omega}^{D}_{\bb,z^{+}}$
that is sufficiently close to the central path,
satisfying $\tilde{\eta}(\ss^{+},s_{gap}^{+}) \leq \frac{1}{10}$.

We prove this formally:
\begin{lemma}[compare {\cite[Lem 4.5]{YeBook}}]\label{lem:shift}
Given $z < z^{*}$ and 
$(\yy , \ss , s_{gap}) \in \interior{\Omega}^{D}_{\bb,z}$ where
$\tilde{\eta}(\ss,s_{gap}) \leq \frac{1}{10}$,
let $s_{gap}' = \bb^{T}\yy - z^{+}$
and $(\yy^{+},\ss^{+},s_{gap}^{+},z^{+}) = \mathtt{Shift}(\yy,\ss,s_{gap},z)$.
Then
\begin{enumerate}[label=(\roman{*}),ref=\ref{lem:shift}(\roman{*})]
\item $z^{+} < z^{*}$
\item \label{lem:shiftprime}
$s_{gap}' > 0$ and $\tilde{\eta}(\ss ,s_{gap}') < \frac{21}{100}$
\item \label{lem:shiftplus}
$\ss^{+},s_{gap}^{+} > 0$ and 
$\tilde{\eta}(\ss^{+} ,s_{gap}^{+}) < \frac{1}{10}$
\end{enumerate}

\end{lemma}
\begin{proof}
(i)
\[
z^{+} 
= 
z + \frac{\bb^{T}\yy - z}{10\sqrt{m}}
<
z + (\bb^{T}\yy - z)
=
\bb^{T}\yy 
<
z^{*}
\]

\medskip\noindent
(ii) 
We note that
$s'_{gap} = s_{gap} - (z^{+} - z) = 
\left(1 - \frac{1}{10\sqrt{m}} \right)s_{gap} > 0$.

Let us write $\tilde{\ss} = \lmx \ss \\ s_{gap}\one_{m} \rmx$
and $\tilde{\ss}' = \lmx \ss \\ s_{gap}'\one_{m} \rmx$
and note that
\begin{equation}\label{eq:sdiff}
\tilde{\ss} - \tilde{\ss}'
=
\lmx 0 \\ (s_{gap}-s'_{gap})\one_{m} \rmx 
= 
\lmx 0 \\ (z^{+} - z)\one_{m} \rmx 
=
\lmx 0 \\ \frac{s_{gap}}{10\sqrt{m}}\one_{m} \rmx 
\end{equation}

Let us define
$\tilde{\xx} = \lmx \xx \\ \xx_{gap} \rmx = 
\xx_{\tilde{A}}(\tilde{\ss})$.
So we have
\begin{align}
\tilde{\eta}(\ss , s_{gap}')
=
\eta_{\tilde{A}}(\tilde{\ss}')
&\leq 
\norm{\tilde{S}'\tilde{\xx} -\one_{2m}} 
\qquad \text{(by Lemma \ref{lem:etamin})}
\notag \\
&\leq 
\norm{\tilde{S}\tilde{\xx} - \one_{2m}} + 
\norm{(\tilde{S}' - \tilde{S})\tilde{\xx}}
\notag \\
&= 
\norm{\tilde{S}\tilde{\xx} - \one_{2m}} + 
\frac{1}{10\sqrt{m}}\norm{s_{gap}\xx_{gap}}
\qquad \text{(by Equation \ref{eq:sdiff})} 
\notag \\
&\leq 
\norm{\tilde{S}\tilde{\xx} - \one_{2m}} + 
\frac{1}{10\sqrt{m}}
\left(\norm{s_{gap}\xx_{gap} - \one_{m}} + \norm{\one_{m}}\right)
\notag \\
&=
\norm{\tilde{S}\tilde{\xx} - \one_{2m}} + 
\frac{1}{10\sqrt{m}}
\norm{s_{gap}\xx_{gap} - \one_{m}} + \frac{1}{10}
\notag \\
&\leq 
\norm{\tilde{S}\tilde{\xx} - \one_{2m}} + 
\frac{1}{10}\norm{s_{gap}\xx_{gap} - \one_{m}} + \frac{1}{10}
\notag \\
&\leq 
\norm{\tilde{S}\tilde{\xx} - \one_{2m}} + 
\frac{1}{10}\norm{\tilde{S}\tilde{\xx} - \one_{2m}} + \frac{1}{10}
\notag \\
&= \frac{11}{10}\tilde{\eta}(\ss,s_{gap}) + \frac{1}{10}
\label{eq:shiftaway} \\
&\leq \frac{11}{10}\cdot \frac{1}{10} + \frac{1}{10} = \frac{21}{100}
\notag
\end{align}

\medskip\noindent
(iii)
By Lemma \ref{lem:newtonstep}, we have $\ss^{+},s_{gap}^{+} > 0$ and
\[
\tilde{\eta}(\ss^{+},s_{gap}^{+}) 
\leq 
\tilde{\eta}(\ss,s_{gap}')^{2} + 
\frac{1}{20}\tilde{\eta}(\ss,s_{gap}')
\leq 
\left(\frac{21}{100}\right)^{2} + \frac{1}{20}\cdot\frac{21}{100} < \frac{1}{10}
\]
\end{proof}

\begin{figure}[t]
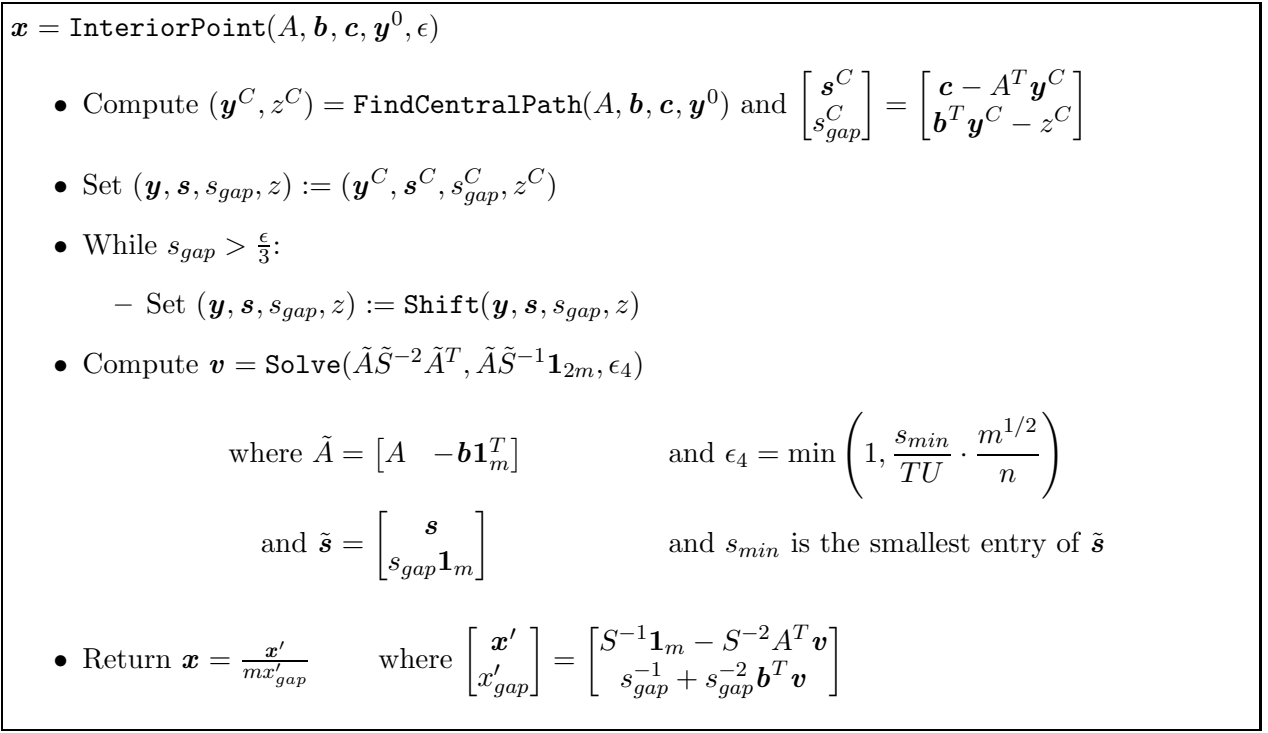

\mybox{
$\xx = \mathtt{InteriorPoint}(A,\bb,\cc,\yy^{0},\epsilon)$
\begin{itemize}
\item Compute
$(\yy^{C},z^{C})=\mathtt{FindCentralPath}(A,\bb,\cc,\yy^{0})$
and
$\lmx \ss^{C} \\ s_{gap}^{C} \rmx =
\lmx \cc - A^{T}\yy^{C} \\ \bb^{T}\yy^{C} - z^{C} \rmx
$
\item Set $(\yy, \ss, s_{gap}, z) := (\yy^{C}, \ss^{C}, s_{gap}^{C}, z^{C})$
\item While $s_{gap} > \frac{\epsilon}{3}$:
\begin{itemize}
\item Set $(\yy,\ss,s_{gap},z) := \mathtt{Shift}(\yy,\ss,s_{gap},z)$
\end{itemize}
\item 
Compute 
$\vv = \mathtt{Solve}(\tilde{A}\tilde{S}^{-2}\tilde{A}^{T},\tilde{A}\tilde{S}^{-1}\one_{2m},\epsilon_{4})$
\begin{align*}
\text{where } \tilde{A} &= \lmx A & -\bb \one_{m}^{T} \rmx
&
&\text{and } 
\epsilon_{4} = \min\left(1, \frac{s_{min}}{TU}\cdot \frac{m^{1/2}}{n} \right)
\\
\text{and } \tilde{\ss} &= \lmx \ss \\ s_{gap}\one_{m} \rmx
&
&\text{and $s_{min}$ }\text{is the smallest entry of $\tilde{\ss}$}
\end{align*}
\item Return 
$\xx = 
\frac{\xx'}
{mx_{gap}'}$
\qquad 
where
$
\lmx \xx' \\ x'_{gap} \rmx =
\lmx S^{-1}\one_{m} - S^{-2}A^{T}\vv \\
s_{gap}^{-1} + s_{gap}^{-2} \bb^{T}\vv \rmx 
$
\end{itemize}
}
\caption{Dual path-following interior-point algorithm
using an approximate solver}
\label{fig:intptalg}
\end{figure}

We now present the complete path-following $\mathtt{InteriorPoint}$ algorithm,
implemented using an approximate solver,
in Figure \ref{fig:intptalg}.
For now we postpone describing
the $\mathtt{FindCentralPath}$ subroutine,
which gives an initial point near the central path.
In particular, it 
produces a $z^{C}<z^{*}$ and $(\yy^{C} ,\ss^{C},s_{gap}^{C}) \in \Omega^{D}_{\bb,z^{C}}$
satisfying $\tilde{\eta}(\ss^{C},s_{gap}^{C}) \leq \frac{1}{10}$.
Once we have this initial central path point,
Lemma \ref{lem:shift} tells us that after each
call to $\mathtt{Shift}$ we have a new value $z < z^{*}$
and new central path point $(\yy,\ss,s_{gap})\in \interior{\Omega}^{D}_{\bb,z}$
that satisfies $\tilde{\eta}(\ss,s_{gap}) \leq \frac{1}{10}$.

Later we will analyze the number of calls to $\mathtt{Shift}$
before the algortihm terminates.
First, let us 
confirm that the algorithm returns the correct output:
\begin{lemma}\label{lem:intptoutput}
The output of $\xx = \mathtt{InteriorPoint}(A,\bb,\cc,\yy^{0},\epsilon)$
satisfies
\begin{itemize}
\item [(i)] $\xx > 0$
\item [(ii)] $\norm{A\xx - \bb} \leq \frac{\epsilon}{12\sqrt{2}\cdot Tn^{1/2}}$
\item [(iii)] $\cc^{T}\xx <  z^{*} + \epsilon$
\end{itemize}
\end{lemma}
\begin{proof}
(i)
To assist in our proof, let us define 
$\tilde{\xx}' = \lmx \xx' \\ x'_{gap}\one_{m} \rmx$ and note that
$\tilde{\xx}' = \tilde{S}^{-1}\one_{2m} - \tilde{S}^{-2}\tilde{A}^{T}\vv$.
We have
\begin{align}
\norm{\tilde{S}\tilde{\xx}' - \one_{2m}}
&=
\norm{\tilde{S}^{-1}\tilde{A}^{T}\vv} 
\notag \\
&=
\norm{\vv}_{\tilde{A}\tilde{S}^{-2}\tilde{A}^{T}} 
\notag \\
&\leq 
\norm{(\tilde{A}\tilde{S}^{-2}\tilde{A}^{T})^{-1}\tilde{A}\tilde{S}^{-1}\one_{2m}}_{\tilde{A}\tilde{S}^{-2}\tilde{A}^{T}} + 
\norm{\vv - (\tilde{A}\tilde{S}^{-2}\tilde{A}^{T})^{-1}\tilde{A}\tilde{S}^{-1}\one_{2m}}_{\tilde{A}\tilde{S}^{-2}\tilde{A}^{T}}  
\notag \\
&\leq 
(1+\epsilon_{4})
\norm{(\tilde{A}\tilde{S}^{-2}\tilde{A}^{T})^{-1}\tilde{A}\tilde{S}^{-1}\one_{2m}}_{\tilde{A}\tilde{S}^{-2}\tilde{A}^{T}} 
\qquad \text{(by guarantee of \texttt{Solve})}
\notag \\
&=
(1+\epsilon_{4})\cdot \tilde{\eta}(\ss ,s_{gap})
\leq 
2\cdot \tilde{\eta}(\ss ,s_{gap})
\leq 
2\cdot \frac{1}{10}
\leq 
\frac{1}{5}
\label{eq:sxprime}
\end{align}

Since $\tilde{\ss}$ is positive, we conclude that $\tilde{\xx}'$
must also be positive, and so must be $\xx$.

\medskip\noindent 
(ii) We have
\begin{align}
\norm{A\xx - \bb}
&=
\frac{1}{mx_{gap}'}
\norm{A\xx' - mx_{gap}'\bb}
\notag \\
&=
\frac{1}{mx_{gap}'}
\norm{\tilde{A}\tilde{\xx}'}
\notag \\
&=
\frac{1}{mx_{gap}'}
\norm{\tilde{A}\tilde{S}^{-1}\one_{2m} -
\tilde{A}\tilde{S}^{-2}\tilde{A}^{T}\vv}
\notag \\
\intertext{Observe that the largest eigenvalue of the matrix 
$\tilde{A}\tilde{A}^{T} = AA^{T} + m\bb \bb^{T}$ 
is less than the trace, which is 
at most $2nmU^{2}$.  Thus, the largest eigenvalue
of $\tilde{A}\tilde{S}^{-2}\tilde{A}^{T}$
is at most $2nmU^{2}s_{min}^{-2}$.  So we proceed
}
&\leq 
\frac{(2nmU^{2}s_{min}^{-2})^{1/2}}{mx_{gap}'}
\norm{\tilde{A}\tilde{S}^{-1}\one_{2m} -
\tilde{A}\tilde{S}^{-2}\tilde{A}^{T}\vv}
_{(\tilde{A}\tilde{S}^{-2}\tilde{A}^{T})^{-1}}
\notag \\
&=
\frac{(2n)^{1/2}U}{m^{1/2}x_{gap}'s_{min}}
\norm{(\tilde{A}\tilde{S}^{-2}\tilde{A}^{T})^{-1}\tilde{A}\tilde{S}^{-1}\one_{2m} - \vv}
_{\tilde{A}\tilde{S}^{-2}\tilde{A}^{T}}
\notag \\
&\leq 
\frac{(2n)^{1/2}U}{m^{1/2}x_{gap}'s_{min}}
\cdot \epsilon_{4} 
\norm{(\tilde{A}\tilde{S}^{-2}\tilde{A}^{T})^{-1}
\tilde{A}\tilde{S}^{-1}\one_{2m}}
_{\tilde{A}\tilde{S}^{-2}\tilde{A}^{T}}
\qquad \text{(by guarantee of \texttt{Solve})}
\notag \\
&=
\frac{(2n)^{1/2}U}{m^{1/2}x_{gap}'s_{min}}
\cdot \epsilon_{4} 
\cdot \tilde{\eta}(\ss,s_{gap})
\notag \\
&\leq 
\frac{(2n)^{1/2}U}{m^{1/2}x_{gap}'s_{min}}
\cdot \epsilon_{4} 
\cdot \frac{1}{10}
\notag \\
&\leq  
\frac{(2n)^{1/2}U}{m^{1/2}x_{gap}'s_{min}}
\cdot 
\frac{m^{1/2}s_{min}}{nTU}
\cdot \frac{1}{10}
\notag \\
&=
\frac{1}{5\sqrt{2}\cdot Tn^{1/2}}\cdot \frac{1}{x_{gap}'}
\notag \\
&\leq 
\frac{1}{5\sqrt{2}\cdot Tn^{1/2}}\cdot \frac{5}{4}\cdot s_{gap}
\qquad \text{(from equation \ref{eq:sxprime}, 
we know $s_{gap}x_{gap}' \geq \frac{4}{5}$)}
\notag \\
&\leq 
\frac{1}{5\sqrt{2}\cdot Tn^{1/2}}\cdot \frac{5}{4}\cdot \frac{\epsilon}{3}
\notag \\
&\leq 
\frac{\epsilon}{12\sqrt{2}\cdot Tn^{1/2}}
\label{eq:axminusb}
\end{align}

\medskip \noindent 
(iii) We have
\begin{align}
\ss^{T}\xx 
&= 
\frac{\ss^{T}\xx '}{mx'_{gap}} 
\notag \\
&\geq 
\frac{5\ss^{T}\xx '}{6m}\cdot s_{gap}
\qquad \text{(from equation \ref{eq:sxprime}, 
we know $s_{gap}x_{gap}' \leq \frac{6}{5}$)}
\notag \\
&=
\frac{5\norm{S\xx'}_{1}}{6m}\cdot s_{gap}
\notag \\
&\geq 
\frac{5}{6m}\left(\norm{\one_{m}}_{1} - \norm{S\xx' - \one_{m}}_{1} \right)\cdot s_{gap}
\notag \\
&=
\frac{5}{6m}\left(m - \norm{S\xx' - \one_{m}}_{1} \right)\cdot s_{gap}
\notag \\
&\geq 
\frac{5}{6m}\left(m - \sqrt{m}\norm{S\xx' - \one_{m}} \right)\cdot s_{gap}
\notag \\
&\geq 
\frac{5}{6m}\left(m - \frac{1}{5}\sqrt{m} \right)\cdot s_{gap}
\qquad \text{(by equation \ref{eq:sxprime})}
\notag \\
&\geq 
\frac{5}{6m}\cdot \frac{4}{5}\cdot  m\cdot s_{gap}
\notag \\
&=
\frac{2}{3}\cdot s_{gap}
\label{eq:sxlowerbound}
\end{align}

\begin{align}
\ss^{T}\xx 
&= 
\frac{\ss^{T}\xx '}{mx'_{gap}} 
\notag \\
&\leq 
\frac{5\ss^{T}\xx '}{4m}\cdot s_{gap}
\qquad \text{(from equation \ref{eq:sxprime}, 
we know $s_{gap}x_{gap}' \geq \frac{4}{5}$)}
\notag \\
&=
\frac{5\norm{S\xx'}_{1}}{4m}\cdot s_{gap}
\notag \\
&\leq 
\frac{5}{4m}\left(\norm{S\xx' - \one_{m}}_{1} + \norm{\one_{m}}_{1} \right)\cdot s_{gap}
\notag \\
&=
\frac{5}{4m}\left(\norm{S\xx' - \one_{m}}_{1} + m \right)\cdot s_{gap}
\notag \\
&\leq 
\frac{5}{4m}\left(\sqrt{m}\norm{S\xx' - \one_{m}} + m \right)\cdot s_{gap}
\notag \\
&\leq 
\frac{5}{4m}\left(\frac{1}{5}\sqrt{m} + m \right)\cdot s_{gap}
\qquad \text{(by equation \ref{eq:sxprime})}
\notag \\
&\leq 
\frac{5}{4m}\cdot \frac{6}{5}\cdot  m\cdot s_{gap}
\notag \\
&=
\frac{3}{2}\cdot s_{gap}
\label{eq:sxupperbound}
\end{align}
We then have
\begin{align*}
\cc^{T}\xx - z^{*}
&>
\cc^{T}\xx - \bb^{T}\yy 
\\
&=
(\cc^{T} - \yy^{T}A)\xx + \yy^{T}(A\xx - \bb)
\\
&=
\ss^{T}\xx + \yy^{T}(A\xx - \bb)
&\geq 
\frac{2}{3}\cdot s_{gap} + \yy^{T}(A\xx - \bb) 
\qquad \text{(by Equation \ref{eq:sxupperbound})}\\
&\leq 
\dots 
\frac{2}{3}\cdot \epsilon + \yy^{T}(A\xx - \bb) \\
&\leq 
\frac{2}{3}\cdot \epsilon + \norm{\yy}\norm{A\xx - \bb} \\
&\leq 
\frac{2}{3}\cdot \epsilon + Tn^{1/2}\norm{A\xx - \bb} \\
&\leq 
\frac{2}{3}\cdot \epsilon + \frac{1}{12\sqrt{2}}\cdot \epsilon 
\qquad \text{(by Lemma \ref{lem:intptoutput}(ii))}\\
&<
\epsilon 
\end{align*}
\begin{align*}
\cc^{T}\xx - z^{*} 
&<
\cc^{T}\xx - z \\
&=
(\cc^{T} - \yy^{T}A)\xx + \yy^{T}(A\xx - \bb) + \bb^{T}\yy - z \\
&=
\ss^{T}\xx + \yy^{T}(A\xx - \bb) + s_{gap} \\
&\leq 
\frac{5}{2}\cdot s_{gap} + \yy^{T}(A\xx - \bb) 
\qquad \text{(by Equation \ref{eq:sxupperbound})}\\
&\leq 
\frac{5}{6}\cdot \epsilon + \yy^{T}(A\xx - \bb) \\
&\leq 
\frac{5}{6}\cdot \epsilon + \norm{\yy}\norm{A\xx - \bb} \\
&\leq 
\frac{5}{6}\cdot \epsilon + Tn^{1/2}\norm{A\xx - \bb} \\
&\leq 
\frac{5}{6}\cdot \epsilon + \frac{1}{12\sqrt{2}}\cdot \epsilon 
\qquad \text{(by Lemma \ref{lem:intptoutput}(ii))}\\
&<
\epsilon 
\end{align*}
\end{proof}

Next, we analyze the number of $\mathtt{Shift}$ iterations
until the algorithm terminates.
We can measure the progress of the algorithm with the potential function
$B(z)$:
\[
B(z) = \sum_{j=1}^{m} \log \overstar{\ss}_{j} + m\log \overstar{s}_{gap}
\qquad \text{where $(\overstar{\yy},\overstar{\ss},\overstar{s}_{gap})$ is the analytic center of $\Omega^{D}_{\bb,z}$}
\]

Soon, we will show how 
a decrease in $B(z)$ implies 
that $s_{gap}$ is decreasing and thus the algorithm is making progress.
Let us first show 
that the value of $B(z)$ decreases by $\Omega(\sqrt{m})$ after
each iteration.

\begin{lemma}[compare {\cite[Lem 4.6]{YeBook}}]\label{lem:decpot}
Given $(\yy,\ss,s_{gap})\in \interior{\Omega}^{D}_{\bb,z}$
satisfiying $\tilde{\eta}(\ss,s_{gap}) < \frac{1}{10}$,
let $(\yy^{+},z^{+}) = 
\mathtt{Shift}(\yy,z)$.
Then $B(z^{+}) \leq B(z) - \Theta (\sqrt{m})$.
\end{lemma}
\begin{proof}
Let $(\overstar{\yy},\overstar{\ss},\overstar{s}_{gap})$ and $(\overstar{\yy}^{+},\overstar{\ss}^{+},\overstar{s}_{gap}^{+})$
respectively be the analytic centers of $\Omega^{D}_{\bb,z}$ and $\Omega^{D}_{\bb,z^{+}}$.

Following Lemma \ref{lem:constructx}, we define
$\xx = \frac{\overstar{s}_{gap}}{m}\overstar{S}^{-1}\one_{m}$
that satisfies $A\xx = \bb$.
We have
\begin{align}
e^{\frac{B(z^{+}) - B(z)}{2m}} 
&= 
\sqrt{
\left(
\prod_{j=1}^{m} \frac{\overstar{\ss}^{+}_{j}}{\overstar{\ss}_{j}}
\right)^{\frac{1}{m}}
\cdot 
\frac{\overstar{s}_{gap}^{+}}{\overstar{s}_{gap}}
}
\notag \\
&\leq 
\frac{1}{2m}\sum_{j=1}^{m} 
\frac{\overstar{\ss}^{+}_{j}}{\overstar{\ss}_{j}}
+
\frac{1}{2}
\frac{\overstar{s}_{gap}^{+}}{\overstar{s}_{gap}}
\notag \\
&\leq
1+
\frac{1}{2m}\sum_{j=1}^{m} 
\frac{\overstar{\ss}^{+}_{j}-\overstar{\ss}_{j}}{\overstar{\ss}_{j}}
+
\frac{1}{2}
\frac{\overstar{s}_{gap}^{+}-\overstar{s}_{gap}}{\overstar{s}_{gap}}
\notag \\
&=
1+
\frac{1}{2\overstar{s}_{gap}}\cdot 
\frac{\overstar{s}_{gap}}{m} 
(\overstar{\ss}^{+}-\overstar{\ss})\overstar{S}^{-1}\one_{m}
+
\frac{1}{2\overstar{s}_{gap}}
(\overstar{s}_{gap}^{+}-\overstar{s}_{gap})
\notag \\
&=
1+
\frac{1}{2\overstar{s}_{gap}} 
\left(
(\overstar{\ss}^{+}-\overstar{\ss})^{T}\xx 
+
(\overstar{s}_{gap}^{+} - \overstar{s}_{gap})
\right)
\notag \\
&=
1+
\frac{1}{2\overstar{s}_{gap}} 
\left(
\left((\cc - A^{T}\overstar{\yy}^{+})-(\cc - A^{T}\overstar{\yy}) \right)^{T}\xx 
+
(\overstar{s}_{gap}^{+} - \overstar{s}_{gap})
\right)
\notag \\
&=
1+
\frac{1}{2\overstar{s}_{gap}} 
\left(
(\overstar{\yy} - \overstar{\yy}^{+})^{T}A\xx 
+
(\overstar{s}_{gap}^{+} - \overstar{s}_{gap})
\right)
\notag \\
&=
1+
\frac{1}{2\overstar{s}_{gap}} 
\left(
(\overstar{\yy} - \overstar{\yy}^{+})^{T}\bb 
+
(\overstar{s}_{gap}^{+} - \overstar{s}_{gap})
\right)
\qquad \text{(by Lemma \ref{lem:constructx})}
\notag \\
&=
1+
\frac{1}{2\overstar{s}_{gap}} 
\left(
(\bb^{T}\overstar{\yy} - \overstar{s}_{gap})
-
(\bb^{T}\overstar{\yy}^{+} - \overstar{s}_{gap}^{+})
\right)
\notag \\
&=
1+
\frac{1}{2\overstar{s}_{gap}} 
\left(
z - z^{+}
\right)
\label{eq:middecpot} \\
&=
1-\frac{s_{gap}}{20\sqrt{m}\cdot \overstar{s}_{gap}}
\notag \\
&\leq 
1-\frac{9}{200\sqrt{m}}
\qquad \text{($\frac{\overstar{s}_{gap}}{s_{gap}} \leq \frac{10}{9}$ by Lemma \ref{lem:nearcenter})}
\notag \\
&\leq 
e^{-\frac{9}{200\sqrt{m}}}
\notag
\end{align}

We conclude that $B(z^{+}) - B(z) \leq -\frac{9}{100}\sqrt{m}$.
\end{proof}

Let us now show that a decrease in
the potential function $B(z)$ 
implies a decrease in the value of $s_{gap}$:
\begin{lemma}[compare {\cite[Prop 4.2]{YeBook}}]\label{lem:potprogress}
Given $(\yy,\ss,s_{gap})\in \interior{\Omega}^{D}_{\bb,z}$
and $(\yy^{+},\ss^{+},s_{gap}^{+})\in \interior{\Omega}^{D}_{\bb,z^{+}}$
where $z^{+}>z$ and
$\tilde{\eta}(\ss,s_{gap}) \leq \eta$
and $\tilde{\eta}(\ss^{+},s_{gap}^{+}) \leq \eta$
for $\eta < 1$.
Then
\[
\frac{s_{gap}}{s_{gap}^{+}}
\leq
(1-2\eta)\cdot \left( e^{\frac{B(z_{1})-B(z_{2})}{m}} - 1 \right)
\]
\end{lemma}

\begin{proof}
Let $(\overstar{\yy},\overstar{\ss},\overstar{s}_{gap})$ and $(\overstar{\yy}^{+},\overstar{\ss}^{+},\overstar{s}_{gap}^{+})$
respectively be the analytic centers of $\Omega^{D}_{\bb,z}$ and $\Omega^{D}_{\bb,z^{+}}$.
We define
$\xx = \frac{\overstar{s}_{gap}}{m}\overstar{S}^{-1}\one_{m}$ and
$\xx^{+} = \frac{\overstar{s}_{gap}^{+}}{m}(\overstar{S}^{+})^{-1}\one_{m}$,
which by Lemma \ref{lem:constructx} satisfy $A\xx = \bb = A\xx^{+}$.

We have
\begin{align*}
e^{\frac{B(z)-B(z^{+})}{m}}
&=
\left( \prod_{j=1}^{m}\frac{\overstar{\ss}_{j}}{\overstar{\ss}^{+}_{j}} \right)^{\frac{1}{m}}
\cdot 
\frac{\overstar{s}_{gap}}{\overstar{s}_{gap}^{+}}
\\
&\leq 
\frac{1}{m}\left( \sum_{j=1}^{m}\frac{\overstar{\ss}_{j}}{\overstar{\ss}^{+}_{j}} \right)
\cdot 
\frac{\overstar{s}_{gap}}{\overstar{s}_{gap}^{+}}
\\
&=
\left(1+ \frac{1}{m}\sum_{j=1}^{m}\frac{\overstar{\ss}_{j}-\overstar{\ss}^{+}_{j}}{\overstar{\ss}^{+}_{j}} \right)
\cdot 
\frac{\overstar{s}_{gap}}{\overstar{s}_{gap}^{+}}
\\
&=
\left(1+ \frac{1}{\overstar{s}_{gap}^{+}}(\overstar{\ss}-\overstar{\ss}^{+})^{T}\xx^{+} \right)
\cdot 
\frac{\overstar{s}_{gap}}{\overstar{s}_{gap}^{+}}
\\
&=
\left(1+ \frac{1}{\overstar{s}_{gap}^{+}}\left((\cc - A^{T}\overstar{\yy})-(\cc - A^{T}\overstar{\yy}^{+})\right)^{T}\xx^{+} \right)
\cdot 
\frac{\overstar{s}_{gap}}{\overstar{s}_{gap}^{+}}
\\
&=
\left(1+ \frac{1}{\overstar{s}_{gap}^{+}}(\overstar{\yy}-\overstar{\yy}^{+})^{T}A\xx^{+} \right)
\cdot 
\frac{\overstar{s}_{gap}}{\overstar{s}_{gap}^{+}}
\\
&=
\left(1+ \frac{1}{\overstar{s}_{gap}^{+}}(\overstar{\yy}-\overstar{\yy}^{+})^{T}\bb \right)
\cdot 
\frac{\overstar{s}_{gap}}{\overstar{s}_{gap}^{+}}
\\
&=
\left(1+ \frac{1}{\overstar{s}_{gap}^{+}}(\overstar{\yy}-\overstar{\yy}^{+})^{T}A\xx \right)
\cdot 
\frac{\overstar{s}_{gap}}{\overstar{s}_{gap}^{+}}
\\
&=
\left(1+ \frac{1}{\overstar{s}_{gap}^{+}}\left((\cc - A^{T}\overstar{\yy})-(\cc - A^{T}\overstar{\yy}^{+})\right)^{T}\xx \right)
\cdot 
\frac{\overstar{s}_{gap}}{\overstar{s}_{gap}^{+}}
\\
&=
\left(1+ \frac{1}{\overstar{s}_{gap}^{+}}(\overstar{\ss}-\overstar{\ss}^{+})^{T}\xx \right)
\cdot 
\frac{\overstar{s}_{gap}}{\overstar{s}_{gap}^{+}}
\\
&\leq 
\left(1+ \frac{1}{\overstar{s}_{gap}^{+}}\overstar{\ss}^{T}\xx \right)
\cdot 
\frac{\overstar{s}_{gap}}{\overstar{s}_{gap}^{+}}
\\
&= 
\left(1 + \frac{\overstar{s}_{gap}}{\overstar{s}_{gap}^{+}} \right)
\cdot 
\frac{\overstar{s}_{gap}}{\overstar{s}_{gap}^{+}}
\\
&\leq  
\left(1 + \frac{\overstar{s}_{gap}}{\overstar{s}_{gap}^{+}} \right)^{2}
\end{align*}

So
\begin{align*}
\frac{\overstar{s}_{gap}}{\overstar{s}_{gap}^{+}}
&\geq 
e^{\frac{B(z)-B(z^{+})}{2m}} - 1
\end{align*}

Using Lemma \ref{lem:nearcenter}, we may conclude
\[
\frac{s_{gap}}{s_{gap}^{+}}
=
\frac{s_{gap}}{\overstar{s}_{gap}}
\cdot 
\frac{\overstar{s}_{gap}}{\overstar{s}_{gap}^{+}}
\cdot 
\frac{\overstar{s}_{gap}^{+}}{s_{gap}^{+}}
\geq 
\frac{1 - \frac{\eta}{1-\eta}}{1 + \frac{\eta}{1-\eta}}\cdot 
\left( e^{\frac{B(z)-B(z^{+})}{2m}} - 1 \right)
\]

\end{proof}

\begin{corollary}\label{cor:shiftcalls}
The $\mathtt{InteriorPoint}$ algorithm makes
$\bigO{\sqrt{m}\log \frac{s_{gap}^{C}}{\epsilon}}$ calls to $\mathtt{Shift}$.
\end{corollary}
\begin{proof}

Recall that the algorithm will terminate only when the value of $s_{gap}$
has decreased from its initial value of $s_{gap}^{C}$ to 
below $\frac{\epsilon}{3}$.
Thus, Lemma \ref{lem:potprogress} ensures us that $s_{gap}$ will be smaller
than $\frac{\epsilon}{3}$ once $B(z)$ has decreased by 
$\Omega \left(m\log \frac{s_{gap}^{C}}{\epsilon}\right)$.
According to Lemma \ref{lem:decpot}, this occurs after 
$\bigO{\sqrt{m}\log \frac{s_{gap}^{C}}{\epsilon}}$ $\mathtt{Shift}$ 
iterations.
\end{proof}

\subsection{Finding the Central Path}

It remains for us to describe how to initialize the path-following algorithm
by finding a point near the central path.
Essentially, this is accomplished by 
running the path-following algorithm in reverse.
Instead of stepping towards the optimum given by $\bb$,
we step away from the optimum given by the vector 
$\hat{\bb} = A(S^{0})^{-1}\one_{m}$ that depends
on our initial feasible point $(\yy^{0} ,\ss^{0}) \in \interior{\Omega}^{D}$.

Our analysis parallels that in the previous section.  
The following function $\hat{\eta}$ measures the proximity
of a point $(\yy,\ss,\hat{s}_{gap})\in 
\interior{\Omega}^{D}_{\hat{\bb},\hat{z}}$
to the central path given by $\hat{\bb}$:
\[
\hat{\eta}(\ss,\hat{s}_{gap}) = \eta_{\tilde{\hat{A}}}(\tilde{\hat{\ss}})
\qquad \text{where $\tilde{\hat{A}} = \lmx A & -\hat{\bb} \one_{m}^{T} \rmx$
and $\tilde{\hat{\ss}} = \lmx \ss \\ \hat{s}_{gap}\one_{m} \rmx $}
\]
To initialize the algorithm, we observe that 
$(\yy^{0},\ss^{0},m)\in \Omega^{D}_{\hat{\bb},\hat{z}^{0}}$ 
is on the $\hat{\bb}$ central path,
where we define $\hat{z}^{0} = \hat{\bb}^{T}\yy^{0} - m$:
\begin{lemma}\label{lem:initpt}
$\hat{\eta}(\ss^{0},m) = 0$
\end{lemma}
\begin{proof}
Defining $\tilde{\hat{\ss}}^{0} = \lmx \ss^{0} \\ m\one_{m} \rmx$, we have
\[
\tilde{\hat{A}}(\tilde{\hat{S}}^{0})^{-1}\one_{2m}
=
\lmx A & -\hat{\bb}\one_{m}^{T} \rmx 
\lmx (S^{0})^{-1}\one_{m} \\ m^{-1}\one_{m} \rmx
=
A(S^{0})^{-1}\one_{m}-  \hat{\bb}\cdot \frac{\one_{m}^{T}\one_{m}}{m}
=
\hat{\bb} - \hat{\bb}
=
0
\]
Thus, 
$\hat{\eta}(\ss^{0},m) = 
\norm{\tilde{\hat{A}}(\tilde{\hat{S}}^{0})^{-1}\one_{2m}}_{(\tilde{\hat{A}}(\tilde{\hat{S}}^{0})^{-2}\tilde{\hat{A}}^{T})^{-1}} = 0$

\end{proof}

\begin{figure}[t]
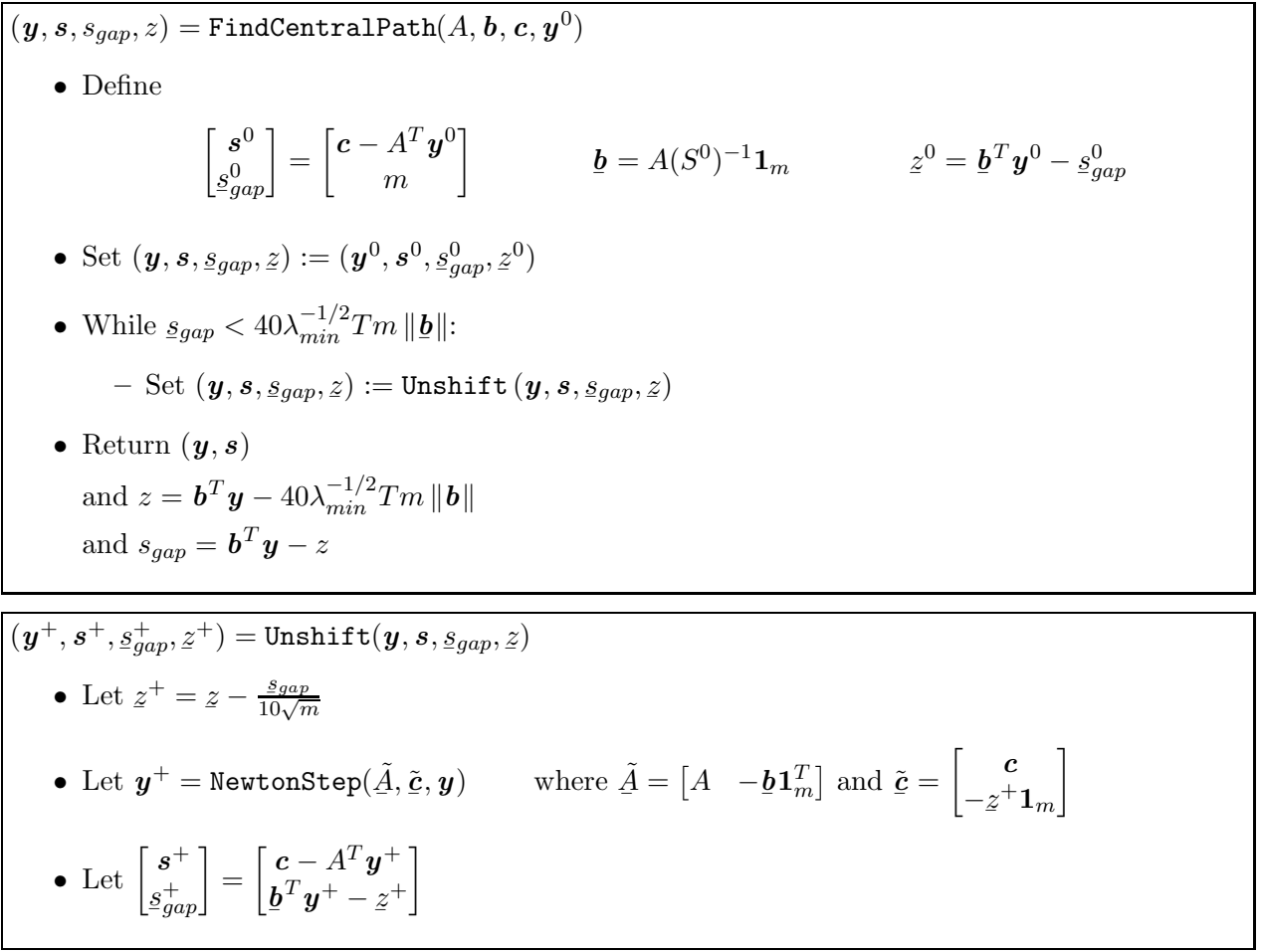

\mybox{
$(\yy,\ss,s_{gap},z) = \mathtt{FindCentralPath}(A,\bb,\cc,\yy^{0})$
\begin{itemize}
\item Define 
\begin{align*}
\lmx \ss^{0} \\ \hat{s}_{gap}^{0} \rmx 
&= 
\lmx \cc - A^{T}\yy^{0} \\ m \rmx 
&
\hat{\bb} 
&= 
A(S^{0})^{-1}\one_{m}
&
\hat{z}^{0} 
&= 
\hat{\bb}^{T}\yy^{0} - \hat{s}_{gap}^{0}
\end{align*}
\item Set $(\yy,\ss,\hat{s}_{gap},\hat{z}):=(\yy^{0},\ss^{0},\hat{s}_{gap}^{0},\hat{z}^{0})$
\item While $\hat{s}_{gap} < 40\lambda_{min}^{-1/2}Tm\norm{\hat{\bb}}$:
\begin{itemize}
\item Set $(\yy,\ss,\hat{s}_{gap},\hat{z}) := 
\mathtt{Unshift}\left(\yy, \ss, \hat{s}_{gap}, \hat{z}\right)$
\end{itemize}
\item Return $(\yy,\ss)$

and
$z = \bb^{T}\yy - 40\lambda_{min}^{-1/2}Tm\norm{\bb}$

and
$s_{gap} = \bb^{T}\yy - z$
\end{itemize}
}
\mybox{
$(\yy^{+},\ss^{+},\hat{s}_{gap}^{+},\hat{z}^{+}) = 
\mathtt{Unshift}(\yy,\ss,\hat{s}_{gap},\hat{z})$
\begin{itemize}
\item Let $\hat{z}^{+} = \hat{z} - \frac{\hat{s}_{gap}}{10\sqrt{m}}$
\item Let $\yy^{+} = 
\mathtt{NewtonStep}(\tilde{\hat{A}},\tilde{\hat{\cc}},\yy)$
\qquad 
where
$\tilde{\hat{A}} = \lmx A & -\hat{\bb}\one_{m}^{T} \rmx$
and
$\tilde{\hat{\cc}} = \lmx \cc \\ -\hat{z}^{+}\one_{m} \rmx$
\item Let $\lmx \ss^{+} \\ \hat{s}_{gap}^{+} \rmx
=
\lmx \cc - A^{T}\yy^{+} \\ \hat{\bb}^{T}\yy^{+} - \hat{z}^{+} \rmx 
$\end{itemize}
}
\caption{Algorithm for finding point near central path given
feasible interior point}
\label{fig:findcentralpath}
\end{figure}

We present the $\mathtt{FindCentralPath}$ algorithm in Figure 
\ref{fig:findcentralpath}.
Starting with $\hat{z} = \hat{z}^{0}$,
we take steps along the $\hat{\bb}$ central path,
decreasing $\hat{z}$ until it is sufficiently small that
the analytic center of 
$\Omega^{D}_{\hat{\bb},\hat{z}}$ 
is close to the analytic center of $\Omega^{D}$,
and therfore also close to the analytic center of $\Omega^{D}_{\bb,z}$
for some sufficiently small $z$.

Let us show that the $\mathtt{Unshift}$ procedure
indeed takes steps near the $\hat{\bb}$ central path:
\begin{lemma}[compare Lemmas \ref{lem:shift}]\label{lem:unshift}
Given $(\yy,\ss,\hat{s}_{gap})\in \interior{\Omega}^{D}_{\hat{\bb},z}$
satisfying $\hat{\eta}(\ss,\hat{s}_{gap}) \leq \frac{1}{40}$.
Let 
$\hat{s}_{gap}' = \bb^{T}\yy - \hat{z}^{+}$
and
$(\yy^{+},\ss^{+},\hat{s}_{gap}^{+},\hat{z}^{+}) = 
\mathtt{Unshift}(\yy,\ss,\hat{s}_{gap},\hat{z})$
.
Then
\begin{enumerate}[label=(\roman{*}),ref=\ref{lem:unshift}(\roman{*})]
\item
$\hat{\eta}(\ss,\hat{s}'_{gap}) \leq \frac{51}{400}$
\item
$\hat{\eta}(\ss^{+},\hat{s}^{+}_{gap}) \leq \frac{1}{40}$.
\end{enumerate}
\end{lemma}
\begin{proof}[Proof of \ref{lem:unshift}(i)]
Following the proof of Lemma \ref{lem:shift}(i)
through equation \ref{eq:shiftaway}, we have
\[
\hat{\eta}(\ss,\hat{s}_{gap}') 
\leq 
\frac{11}{10}\cdot \hat{\eta }(\ss,\hat{s}_{gap}) + \frac{1}{10}
\leq 
\frac{11}{10}\cdot \frac{1}{40} + \frac{1}{10}
=
\frac{51}{400}
\]
\end{proof}
\begin{proof}[Proof of \ref{lem:unshift}(ii)]
By Lemma \ref{lem:newtonstep}, we have
\[
\tilde{\eta}(\ss^{+},\hat{s}_{gap}^{+}) 
\leq 
\tilde{\eta}(\ss,\hat{s}_{gap}')^{2} + 
\frac{1}{20}\tilde{\eta}(\ss,\hat{s}_{gap}')
\leq 
\left(\frac{51}{400}\right)^{2} + \frac{1}{20}\cdot\frac{51}{400} < \frac{1}{40}
\]
\end{proof}

Next, let us prove that the point returned by
$\mathtt{FindCentralPath}$ is
indeed near the original central path (i.e. the path given by $\bb$):
\begin{lemma}\label{lem:findcentralpath}
For $\yy^{0}$ satisfying $A^{T}\yy^{0} < \cc$,
let
$(\yy,\ss,s_{gap},z) = \mathtt{FindCentralPath}(A,\bb,\cc,\yy^{0})$.

Then
$(\yy,\ss,s_{gap}) \in \interior{\Omega}^{D}_{\bb,z}$ and
$\tilde{\eta}(\ss,s_{gap}) \leq \frac{1}{10}$.
\end{lemma}
\begin{proof}
Using the values at the end of the algorithm,
we write $\tilde{\ss} = \lmx \ss \\ s_{gap}\one_{m} \rmx$
and $\tilde{\hat{\ss}} = \lmx \ss \\ \hat{s}_{gap}\one_{m} \rmx$.

To begin, we note
\begin{align}
\hat{s}_{gap} 
&\geq 
40\lambda_{min}^{-1/2}Tm\norm{\hat{\bb}}
\notag \\
&= 
40m(T^{-2}\lambda_{min})^{-1/2}\norm{\hat{\bb}}
\notag \\
&\geq 
40m\norm{\hat{\bb}}_{(AS^{-2}A^{T})^{-1}}
\label{eq:unshift-hatsgap}
\end{align}
where the last inequality 
follows because the smallest eigenvalue of $AS^{-2}A^{T}$
is at least $T^{-2}\lambda_{min}$.

Similarly,
\begin{equation}\label{eq:unshift-sgap}
s_{gap} = 40\lambda_{min}^{-1/2}Tm\norm{\bb}
\geq
40m\norm{\bb}_{(AS^{-2}A^{T})^{-1}}
\end{equation}

We have
\begin{align*}
\tilde{\eta}(\ss,s_{gap})
&=
\norm{\tilde{A}\tilde{S}^{-1}\one_{2m}}_{(\tilde{A}\tilde{S}^{-2}\tilde{A}^{T})^{-1}}
\\
&\leq 
\norm{\tilde{A}\tilde{S}^{-1}\one_{2m}}_{(AS^{-2}A^{T})^{-1}}
\\
&\qquad \text{(because 
$\tilde{A}\tilde{S}^{-2}\tilde{A}^{T} - AS^{-2}A^{T} = ms_{gap}^{-2}\bb \bb^{T}$ is positive semidefinite)}
\\
&=
\norm{\tilde{\hat{A}}\tilde{\hat{S}}^{-1}\one_{2m} - ms_{gap}^{-1}\bb + m\hat{s}_{gap}^{-1}\hat{\bb}}_{(AS^{-2}A^{T})^{-1}}
\\
&\leq 
\norm{\tilde{\hat{A}}\tilde{\hat{S}}^{-1}\one_{2m}}_{(AS^{-2}A^{T})^{-1}} 
+
m\hat{s}_{gap}^{-1}\norm{\hat{\bb}}_{(AS^{-2}A^{T})^{-1}}
+
ms_{gap}^{-1}\norm{\bb}_{(AS^{-2}A^{T})^{-1}}
\\
&\leq 
\norm{\tilde{\hat{A}}\tilde{\hat{S}}^{-1}\one_{2m}}_{(AS^{-2}A^{T})^{-1}} 
+
\frac{1}{40}
+
\frac{1}{40}
\qquad \text{(by equations \ref{eq:unshift-hatsgap} and \ref{eq:unshift-sgap})}
\\
&\leq 
\left(1 + m\hat{s}_{gap}^{-1}\norm{\hat{\bb}}_{(AS^{-2}A^{T})^{-1}}\right)^{1/2}
\norm{\tilde{\hat{A}}\tilde{\hat{S}}^{-1}\one_{2m}}_{(\tilde{\hat{A}}\tilde{\hat{S}}^{-2}\tilde{\hat{A}}^{T})^{-1}}
+
\frac{1}{40}
+
\frac{1}{40}
\\
&\qquad \text{(by Lemma \ref{lem:rankoneupdate}, using the fact that 
$\tilde{\hat{A}}\tilde{\hat{S}}^{-2}\tilde{\hat{A}}^{T} - AS^{-2}A^{T} = m\hat{s}_{gap}^{-2}\hat{\bb}\hat{\bb}^{T}$)}
\\
&\leq 
\left(1 + \frac{1}{40}\right)^{1/2}
\norm{\tilde{\hat{A}}\tilde{\hat{S}}^{-1}\one_{2m}}_{(\tilde{\hat{A}}\tilde{\hat{S}}^{-2}\tilde{\hat{A}}^{T})^{-1}}
+
\frac{1}{40}
+
\frac{1}{40}
\qquad \text{(by equation \ref{eq:unshift-hatsgap})}
\\
&=
\left(1 + \frac{1}{40}\right)^{1/2}
\cdot \hat{\eta}(\ss,\hat{s}_{gap})
+
\frac{1}{40}
+
\frac{1}{40}
\\
&\leq 
2\cdot \hat{\eta}(\tilde{\hat{\ss}},\hat{s}_{gap}) + \frac{1}{20}
\\
&\leq 
2\cdot \frac{1}{40} + \frac{1}{20}
\qquad \text{(by Lemma \ref{lem:unshift}(ii))}
\\
&= 
\frac{1}{10}
\end{align*}
\end{proof}

To measure the progress of the $\mathtt{FindCentralPath}$ algorithm, 
we define $\hat{B}(\hat{z})$:
\[
\hat{B}(\hat{z}) = 
\sum_{j=1}^{m} \log \overstar{\ss}_{j} + m\log \overstar{\hat{s}}_{gap}
\qquad \text{where $(\overstar{\yy},\overstar{\ss},\overstar{\hat{s}}_{gap})$ is the analytic center of $\Omega^{D}_{\hat{\bb},\hat{z}}$}
\]

\begin{lemma}[compare Lemma \ref{lem:decpot}]\label{lem:incpot}
Given $(\yy,\ss,\hat{s}_{gap})\in \interior{\Omega}^{D}_{\hat{\bb},z}$
satisfying $\hat{\eta}(\ss,\hat{s}_{gap}) \leq \frac{1}{40}$.
Let 
$(\yy^{+},\ss^{+},\hat{s}_{gap}^{+},\hat{z}^{+}) = 
\mathtt{Unshift}(\yy,\ss,\hat{s}_{gap},\hat{z})$.
Then
$\hat{B}(\hat{z}^{+}) \geq \hat{B}(\hat{z}) + \Theta (\sqrt{m})$.
\end{lemma}
\begin{proof}
We will follow the proof of Lemma \ref{lem:decpot}, with some minor changes.

Before we proceed, let us recall the definition
$\hat{z}^{+} = \hat{z} - \frac{\hat{s}_{gap}}{10\sqrt{m}}$
to note that
\begin{equation}\label{eq:stosprime}
\hat{s}_{gap}' = 
\hat{s}_{gap} + \hat{z} - \hat{z}^{+} = 
10\sqrt{m}(\hat{z} - \hat{z}^{+}) + \hat{z} - \hat{z}^{+}
\leq 
11\sqrt{m}(\hat{z} - \hat{z}^{+})
\end{equation}

Now, we switch the places of $z$ and $z^{+}$,
and follow the proof of Lemma \ref{lem:decpot} up to 
Equation \ref{eq:middecpot}:
\begin{align*}
e^{\frac{\hat{B}(\hat{z}) - \hat{B}(\hat{z}^{+})}{2m}}
&\leq 
1+\frac{1}{2\overstar{\hat{s}}_{gap}^{+}}\cdot (\hat{z}^{+}-\hat{z})
\\
\intertext{We continue:}
&\leq 
1-\frac{\hat{s}_{gap}'}{22\sqrt{m}\cdot \overstar{\hat{s}}_{gap}^{+}}
\qquad \text{(by Equation \ref{eq:stosprime})}
\\
&\leq 
1 - 
\frac{1}{22\sqrt{m}}\cdot \frac{349}{400}
\qquad \text{(by Lemmas \ref{lem:unshift} and \ref{lem:nearcenter})}
\\
&\leq 
e^{-\frac{169}{4400\sqrt{m}}}
\end{align*}
\end{proof}

\begin{corollary}\label{cor:unshiftcalls}
The $\mathtt{FindCentralPath}$ algorithm makes
$O\left(\sqrt{m}\log \frac{TUm}{\lambda_{min}s^{0}_{min}} \right)$ calls to $\mathtt{Unshift}$,
where $s^{0}_{min}$ is the smallest entry of $\ss^{0} = \cc - A^{T}\yy^{0}$.
\end{corollary}
\begin{proof}
Recall that the algorithm will terminate only when the value of $\hat{s}_{gap}$
has increased from its initial value of $m$ to at least
$40\lambda_{min}^{-1/2}Tm\norm{\hat{\bb}}$.  
So, by Lemma \ref{lem:potprogress}, 
this will have happened once $\hat{B}(z)$ has increased by 
$\Omega \left(m\log \left(\lambda_{min}^{-1/2}T\norm{\hat{\bb}}\right)\right)$
.

According to Lemma \ref{lem:incpot}, this occurs after 
$O \left(\sqrt{m}\log \left(\lambda_{min}^{-1/2}T\norm{\hat{\bb}}\right)\right)$
iterations.

To complete the proof, we note that
\[
\norm{\hat{\bb}} 
= 
\norm{A(S^{0})^{-1}\one_{m}}
\leq 
\frac{n^{1/2}mU}{s^{0}_{min}}
\]
\end{proof}

\subsection{Calls to the Solver}

In each call to $\mathtt{Unshift}$, we solve one
system in a matrix of the form
\[
\tilde{\hat{A}}\tilde{\hat{S}}^{-2}\tilde{\hat{A}}^{T}
=
AS^{-2}A^{T} + m\hat{s}_{gap}^{-2}\hat{\bb}\hat{\bb}^{T}
\]
and in each call to $\mathtt{Shift}$, we solve one
system in a matrix of the form
\[
\tilde{A}\tilde{S}^{-2}\tilde{A}^{T}
=
AS^{-2}A^{T} + ms_{gap}^{-2}\bb\bb^{T}
\]
At the end of the interior-point algorithm we have one final call
of the latter form.

In order to say something about the condition number of the above matrices,
we must bound the slack vector $\ss$.
We are given an upper bound of $T$ on the elements of $\ss$,
so it remains to prove a lower bound:

\begin{lemma}\label{lem:slackslowerbound}
Throughout the $\mathtt{InteriorPoint}$ algorithm,
$\ss \geq \frac{\epsilon}{48nmTU}\ss^{0}$
\end{lemma}
\begin{proof}
At all times duing the algorithm,
we know from Lemma \ref{lem:nearcenter}
that the elements of $\ss$ are bounded by a constant factor from
the slacks at the current central path point $\overstar{\ss}$.
In particular, taking into account 
Lemmas \ref{lem:shift} and \ref{lem:unshift},
we surely have $\ss \geq \frac{1}{2}\overstar{\ss}$.
So let us bound from below the elements of $\overstar{\ss}$.

During the $\mathtt{FindCentralPath}$ subroutine,
as we decrease $\hat{z}$ and expand the polytope 
$\Omega^{D}_{\hat{\bb},\hat{z}}$,
clearly the initial point $\ss^{0}$ remains
in the interior of $\Omega^{D}_{\hat{\bb},\hat{z}}$ throughout.
Thus, by Lemma \ref{lem:boundedslacks},
we have $\overstar{\ss} \geq \frac{1}{2m}\ss^{0}$,
and so $\ss \geq \frac{1}{2}\overstar{\ss} \geq \frac{1}{4m}\ss^{0}$.

Unfortunately,
during the main part of the algorithm, 
as we increase $z$ and shrink the polytope $\Omega^{D}_{\bb,z}$,
the initial point may not remain inside the polytope.
In particular, once we have $z \geq \bb^{T}\yy^{0}$,
the initial point is no longer in $\Omega^{D}_{\bb,z}$,
but we may define a related point $(\yy^{z},\ss^{z},s_{gap}^{z})$
that is in $\Omega^{D}_{\bb,z}$.

Given our current point $(\yy,\ss,s_{gap})\in \Omega^{D}_{\bb,z}$
for $z \geq \bb^{T}\yy^{0}$,
let us define $r = \frac{\bb^{T}\yy-z}{2(\bb^{T}\yy-\bb^{T}\yy^{0})}$
and note that $0 < r < \frac{1}{2}$.  We then define
\begin{align*}
\yy^{z} 
&= 
r \yy^{0} + (1-r)\yy
&
\lmx \ss^{z} \\ s_{gap}^{z} \rmx 
&=
\lmx \cc - A^{T}\yy^{z} \\ \bb^{T}\yy^{z} - z \rmx 
=
\lmx r\ss^{0} + (1-r)\ss \\ \frac{1}{2}(\bb^{T}\yy - z) \rmx > 0
\end{align*}
Therefore Lemma \ref{lem:boundedslacks} gives
\[
\overstar{\ss}
\geq 
\frac{1}{2m}\ss^{z}
=
\frac{r\ss^{0} + (1-r)\ss}{2m}
\geq 
\frac{r}{2m}\ss^{0}
\]

We then find
\[
r 
= 
\frac{s_{gap}}{2(\bb^{T}\yy - \bb^{T}\yy^{0})}
\geq  
\frac{s_{gap}}{4nTU}
\geq  
\frac{\epsilon}{24nTU}
\]
The last inequality follows because, when $s_{gap}$ decreased
below $\frac{\epsilon}{3}$ on the final step,
using Lemma \ref{lem:nearcenter} we find that
it certainly could not have
decreased by more than a factor of $\frac{1}{2}$.

We conclude $\ss \geq \frac{1}{2}\overstar{\ss} \geq 
\frac{r}{4m}\ss^{0} \geq \frac{\epsilon}{48nmTU}\ss^{0}$
\end{proof}

We may now summarize the calls to the solver as follows:
\begin{theorem}
The $\mathtt{InteriorPoint}(A,\bb,\cc,\yy^{0},\epsilon)$ algorithm
makes $\bigO{\sqrt{m}\log \frac{TUm}{\lambda_{min}s^{0}_{min}\epsilon}}$
calls to the approximate solver, of the form
\[
\mathtt{Solve}\left(AS^{-2}A^{T} + \vv\vv^{T},\cdot,\Theta(m^{-1/2})\right)
\]
and one call of the form
\[
\mathtt{Solve}\left(AS^{-2}A^{T} + \vv\vv^{T},\cdot,
\Omega\left(\frac{s^{0}_{min}\epsilon}{m^{1/2}n^{2}T^{2}U^{2}}\right)\right)
\]
where $S$ is a positive diagonal matrix with condition number
$\bigO{\frac{T^{2}Um^{2}}{\epsilon}}$, and $\vv$ satisfies
\[
\norm{\vv} 
= 
\bigO{\frac{U(mn)^{1/2}}{s^{0}_{min}\epsilon}}
\]
\end{theorem}

\begin{proof}
From Lemmas \ref{cor:unshiftcalls} and \ref{cor:shiftcalls},
the total number of solves is
$\bigO{\sqrt{m}\left(
\log \frac{TUm}{\lambda_{min}s^{0}_{min}} +  
\log \frac{s_{gap}^{C}}{\epsilon}
\right)}$,
where we know from the $\mathtt{FindCentralPath}$ algorithm
that $s_{gap}^{C} = 40\frac{Tm\norm{\bb}}{\lambda_{min}^{1/2}}
= \bigO{\frac{TUmn^{1/2}}{\lambda_{min}^{1/2}}}$

As we noted above,
all solves are in matrices that take the form $AS^{-2}A^{T} + \vv\vv^{T}$,
where 
\begin{align*}
\vv 
&= 
m^{1/2}s_{gap}^{-1}\bb
&
\text{or}
&&
\vv 
&= 
m^{1/2}\hat{s}_{gap}^{-1}A(S^{0})^{-1}\one_{m}
\\
\intertext{
We know that $s_{gap} = \Omega(\epsilon)$ and $\hat{s}_{gap} = \Omega(m)$,
so we obtain the respective bounds
}
\norm{\vv}
&=
\bigO{\frac{U(mn)^{1/2}}{\epsilon}}
&&&
\norm{\vv}
&=
\bigO{\frac{U(mn)^{1/2}}{s^{0}_{min}}}
\end{align*}

The condition number of $S$ comes from Lemma \ref{lem:slackslowerbound}
and the upper bound of $T$ on the slacks.

The error parameter for the solver is $\Theta\left(m^{-1/2} \right)$
from the all $\mathtt{NewtonStep}$ calls.
In the final solve, the error parameter is
$\frac{s_{min}m^{1/2}}{TUn} \geq 
\frac{m^{1/2}}{TUn}\cdot \frac{s^{0}_{min}\epsilon}{48nmTU}$,
again invoking Lemma \ref{lem:slackslowerbound}.
\end{proof}


\comment{
\newpage

\section{To do}
Add a conclusion with:
\begin{itemize}
\item [1.] open question - other problems,
\item [2.] ack of impracticality
\item [3.] why only handle lossy
\end{itemize}

\begin{enumerate}
\item [1.] Put upper bound on slacks into complexity bound.  Maybe by
  an absolute upper bound on slacks.
\item [2.] write rounding algorithm to flows
\item [3.] write rounding for ordinary max flow (Dan)

\item [4.] move up linear solver stuff - reduction to m-matrices.

\item [5.] M-matrices in the title?

\item [6.] move approximate solve theorem to intro
\end{enumerate}
}

\end{document}